\documentclass[fleqn,usenatbib]{mnras}

\usepackage[T1]{fontenc}

\DeclareRobustCommand{\VAN}[3]{#2}
\let\VANthebibliography\thebibliography
\def\thebibliography{\DeclareRobustCommand{\VAN}[3]{##3}\VANthebibliography}

\usepackage{graphicx}	
\usepackage{amsmath}	
\usepackage{amssymb}	
\usepackage{comment}
\usepackage{verbatim}
\title[The magnetic field in Abell 2345]{The intra-cluster magnetic field in the double relic galaxy cluster Abell 2345}

\author[Stuardi et al.]{Stuardi, C.,$^{1,2}$\thanks{E-mail: ccstuardi@gmail.com}
Bonafede, A.,$^{1,2}$
Lovisari, L.,$^{3,4}$
Dom\'{i}nguez-Fern\'{a}ndez, P.,$^{5}$
Vazza, F.,$^{1,2,5}$
\newauthor{Br\"uggen, M.,$^{5}$
van Weeren, R.J.,$^{6}$
de Gasperin, F.$^{5}$}
\\
$^{1}$Dipartimento di Fisica e Astronomia, Universit\`a di Bologna, via Gobetti 93/2, I-40129 Bologna, Italy\\
$^{2}$INAF - Istituto di Radioastronomia di Bologna, Via Gobetti 101, I-40129 Bologna, Italy\\
$^{3}$INAF - Osservatorio di Astrofisica e Scienza dello Spazio di Bologna, via Piero Gobetti 93/3, I-40129 Bologna, Italy \\
$^{4}$Center for Astrophysics $|$ Harvard \& Smithsonian, 60 Garden Street, Cambridge, MA 02138, USA \\
$^{5}$Hamburger Sternwarte, Universit\"at Hamburg, Gojenbergsweg 112, 21029 Hamburg, Germany\\
$^{6}$Leiden Observatory, Leiden University, PO Box 9513, 2300 RA Leiden, The Netherlands\\
}

\date{Accepted XXX. Received YYY; in original form ZZZ}

\pubyear{2020}

\begin{document}
\label{firstpage}
\pagerange{\pageref{firstpage}--\pageref{lastpage}}
\maketitle

\begin{abstract}
Magnetic fields are ubiquitous in galaxy clusters, yet their radial profile, power spectrum, and connection to host cluster properties are poorly known. Merging galaxy clusters hosting diffuse polarized emission in the form of radio relics offer a unique possibility to study the magnetic fields in these complex systems.
In this paper, we investigate the intra-cluster magnetic field in Abell 2345. This cluster hosts two radio relics that we detected in polarization with 1-2 GHz JVLA observations. X-ray XMM-Newton images show a very disturbed morphology.
We derived the Rotation Measure (RM) of five polarized sources within $\sim$1 Mpc from the cluster center applying the RM synthesis. Both, the average RM and the RM dispersion radial profiles probe the presence of intra-cluster magnetic fields. Using the thermal electron density profile derived from X-ray analysis and simulating a 3D magnetic field with fluctuations following a power spectrum derived from magneto-hydrodynamical cosmological simulations, we build mock RM images of the cluster. We constrained the magnetic field profile in the eastern radio relic sector by comparing simulated and observed RM images. We find that, within the framework of our model, the data require a magnetic field scaling with thermal electron density as $B(r)\propto n_e(r)$. The best model has a central magnetic field (within a 200 kpc radius) of 2.8$\pm0.1 \ \mu$G. The average magnetic field at the position of the eastern relic is $\sim0.3 \ \mu$G, a factor 2.7 lower than the equipartition estimate.
\end{abstract}

\begin{keywords}
magnetic fields  --  radiation mechanisms: non thermal  --  galaxies: clusters: individual: Abell 2345.
\end{keywords}



\section{Introduction}

Galaxy clusters are unique laboratories for the study of large-scale magnetic fields and their amplification \citep[see][for some reviews]{Ryu12,Donnert18}. The turbulent intra-cluster medium (ICM) is in fact permeated by a ubiquitous magnetic field of 0.1-10 $\mu$G, tangled on scales ranging from few to hundreds of kpc \citep{Bruggen13}. Although the presence of large-scale magnetic fields has been detected beyond doubt, the effective strength, structure and connection to the dynamical state of the clusters are still poorly known.

Large-scale magnetic fields are clearly unveiled by the presence of diffuse cluster radio emission which, in merging galaxy clusters, is detected in the form of radio halos and radio relics \citep[see][for a recent review]{vanWeeren19}. Radio halos are round-shape Mpc-sized sources, centrally located in merging galaxy clusters. Radio relics are Mpc-sized sources that show high levels of polarization ($>10\%$ at GHz frequencies) and are located on the outskirts of merging galaxy clusters. They often have an arc-like shape. Both halos and relics have steep radio spectra ($\alpha>1$, where the flux density is defined as $S_{\nu}\propto\nu^{-\alpha}$) and low surface brightness at GHz frequencies ($\sim$1 $\mu$Jy arcsec$^{-2}$). 

The origin of such diffuse sources is thought to be connected to the formation history of the cluster. In particular, major merger events induce, both, shock waves and turbulence. While the former are associated with the origin of radio relics, the latter is thought to give rise to radio halos \citep[e.g.,][]{Ferrari08,Brunetti14}. The particle acceleration mechanisms involved in these processes are not fully clear, but they must be able to accelerate (or re-accelerate) cosmic-ray electrons up to relativistic energies that emit synchrotron emission in the cluster magnetic field.

Another way in which the presence of cluster magnetic fields is unveiled is the Faraday rotation effect caused by the magneto-ionic ICM on linearly polarized radiation \citep[e.g.,][]{Burn66}. This effect causes the rotation of the polarization angle $\chi$ of polarized sources, seen in projection behind or within galaxy clusters. The rotation is proportional to the squared wavelength of the emission, $\lambda^2$, and to the Faraday depth, $\phi$:
\begin{equation}
   \phi=0.812\int_{\rm{source}}^{\rm{observer}} {n_e B_{\parallel} \text{d}l} \quad {\rm rad \ m^{-2}} \ ,
\label{eq:RM}
\end{equation}
where $n_e$ is the thermal electron density in cm$^{-3}$, $B_{\parallel}$ is the magnetic field component parallel to the line-of-sight in $\mu$G, and d$l$ is the infinitesimal path length in parsecs. A positive Faraday depth implies a magnetic field pointing toward the observer. The Faraday depth coincides with the Rotation Measure (RM) when the rotation is caused by one or several non-emitting Faraday screens \citep{Brentjens05}.

Faraday rotation studies have led to great improvement in our knowledge of cluster magnetic fields \citep[e.g.,][]{Clarke01,Murgia04,Bonafede10,Bohringer16}. In particular, it is now clear that the magnetic field strength decreases with the distance from the cluster center, yet fluctuating over a range of spatial scales. In the most simplistic approach, it can be characterized by its power-spectrum and radial dependence. The magnetic power spectrum is not well known and often assumed to follow the same trend as the velocity power spectrum, i.e., a Kolmogorov spectrum. However, recent cosmological magneto-hydrodynamical (MHD) simulations of galaxy cluster formation have shown that the magnetic spectra arising from the dynamic of the ICM are more complex than power-law spectra \citep{Vazza18,Dominguez19}. This is in agreement with the idealized simulations of magnetic field growth due to the small-scale dynamo \citep{Schekochihin02,Ryu08}. RM observations are fundamental to determine the characteristics of the intra-cluster magnetic fields since the RM of the sources and their dispersion reveal reveal, both, the strength and the structure of the magnetic field along the line-of-sight.

In this paper, we study the magnetic field in the merging galaxy cluster Abell 2345 using Jansky Very Large Array (JVLA) observations in the 1-2 GHz band, XMM-Newton observations, and numerical simulations of the cluster magnetic field. Abell 2345 is highly disturbed and hosts two radio relics. The aim of this work is to constrain the magnetic field profile in the cluster, up to the peripheral regions where the relics are located. Using recent results from MHD cosmological simulations, we produce mock RM images and compare them with observed RM data. This work will improve our understanding on how the RMs derived from relics can be used in order to derive general information on the magnetic fields in the cluster, as well as to constrain the magnetic fields at the relics and its amplification. In Sec.~\ref{sec:analysis} we describe the radio observations, data reduction and imaging techniques, both in continuum and in polarization, and the X-ray data analysis. In Sec.~\ref{sec:reluts}, we show X-ray results and discuss the results of the polarization and RM synthesis analysis. In Sec.~\ref{sec:Bmodel} we describe our simulations and we constrain intra-cluster magnetic field properties. We discuss our results and conclude with a summary in Sec.~\ref{sec:discus} and \ref{sec:conclusion}. 

Throughout this paper, we assume a $\Lambda$CDM cosmological model, with $H_0$ = 69.6 km s$^{-1}$ Mpc$^{-1}$, $\Omega_\text{M}$ = 0.286, $\Omega_{\Lambda}$ = 0.714 \citep{Bennett14}. With this cosmology 1$\arcsec$ corresponds to 3.043 kpc at the cluster redshift, $z$=0.1789.

\subsection{Abell 2345}
\label{sec:A2345}

\begin{table}
	\centering
	\caption{Properties of A2345. Row 1,2: J2000 celestial coordinates  of the X-ray cluster peak; Row 3: redshift; Row 4: X-ray luminosity in the energy band 0.1-2.4 keV; Row 5: estimate of the hydrostatic mass from Sunyaev-Zeldovich effect observation.  References: (1) This work, (2) \citet{Boschin10}, (3) \citet{Lovisari20}, (4) \citet{Planck16b}.}
	\label{tab:info}
	\begin{tabular}{lcc}
		\hline
		R.A. (J2000) & 21$^{\rm h}$27$^{\rm m}$12$^{\rm s}$.6 & (1) \\
		Dec. (J2000) & -12$^{\circ}$09$\arcmin$46$\arcsec$ & (1) \\
		$z$ & 0.1789 & (2) \\
		$L_{\rm X[0.1-2.4 keV]}$ & 2.91$ \times10^{44}$ erg s$^{-1}$ & (3) \\
		$M^{\rm SZ}_{500}$ & 5.91$ \times10^{14}{\rm  M_\odot}$ & (4) \\
		\hline
	\end{tabular}
\end{table}

Abell 2345 (A2345, $z$=0.1789, \citealt{Boschin10}) is a rich galaxy cluster, cataloged as one of the brightest X-ray clusters within the ROentgen SATellite (ROSAT) All Sky Survey \citep{Ebeling96}. The main properties of this cluster are listed in Tab.~\ref{tab:info}. A detailed X-ray study of A2345 is still missing, but several authors pointed out its disturbed morphology as shown by ROSAT, Chandra and XMM-Newton observations \citep[e.g.,][]{Lovisari17,Golovich19a}. \citet{Rossetti16} found a significant offset of $\sim200 \ \rm kpc$ between the X-ray peak of A2345 and its Brightest Cluster Galaxy (BCG, at the J2000 coordinates: 21$^{\rm h}$27$^{\rm m}$13$^{\rm s}$.7, -12$^{\circ}$09$\arcmin$47$\arcsec$), confirming a highly disturbed X-ray morphology.

The presence of diffuse radio emission in the A2345 cluster was discovered by \citet{Giovannini99}. Using images of the National Radio Astronomy Observatory VLA Sky Survey (NVSS), two candidate radio relics were observed in the outskirts of this cluster, on opposite sides with respect to the cluster center: to the east (E relic) and to the west (W relic). A detailed radio analysis of this cluster, including spectral index and polarization analysis, was performed by \citet{Bonafede09a}. The authors used VLA observations at 325 MHz and 1.4 GHz. The W relic revealed a peculiar morphology with a faint filamentary structure extending toward the cluster outskirts. The spectral index image of this radio relic shows a steepening toward the cluster outskirts, opposite to other radio relics, which steepen toward the cluster center \citep[see, e.g.,][]{vanWeeren10}. Together with the comparison with the ROSAT image, this observation suggested that the W relic was produced by a complex merger between different sub-groups. The E relic instead, elongated along the north-south direction, can be more easily explained by a major merger along the main E-W axis. At 1.4 GHz and at the resolution of $23\arcsec\times16\arcsec$ the authors found a mean fractional polarization of the E relic of $\sim 22 \ \%$, reaching values up to $50 \ \%$ in the eastern region. The W relic instead shows a mean fractional polarization of $\sim 14 \ \%$, with regions of higher fractional polarization ($\sim 60 \ \%$) in the northwestern part of the relic. \citet{Bonafede09a} also estimated the equipartition magnetic field in the W and E relic to be 1.0  and 0.8 $\mu G$, respectively. Recently, \citet{George17} computed the integrated spectral indices of the two relics between 118 MHz and 1.4 GHz, obtaining values consistent with the work of \citet{Bonafede09a}: $\alpha$ = 1.29$\pm$0.07 for the E relic and $\alpha$ = 1.52$\pm$0.08 for the W relic. 

\citet{Dahle02} performed a weak lensing analysis of this cluster, considering a small field-of-view of $6\arcmin\times6\arcmin$, centered on the main eastern sub-cluster of A2345. Although they noticed numerous substructures in the ROSAT image, suggestive of a dynamically disturbed system, the weak lensing analysis resulted in a density distribution roughly peaked around the BCG. The main peak in the mass map was found at $\sim1\arcmin.5$ (i.e., 274 kpc) to the east of the BCG. These results were confirmed by \citet{Cypriano04}. A weak lensing study on a larger field-of-view, comprising the entire cluster up to the virial radius, found instead numerous substructures for which it was classified as complex \citep{Okabe10}. In this latter study, a spherically symmetric morphology was discarded.

\citet{Boschin10} performed an extensive optical study of the A2345 cluster to unveil its internal dynamics. The presence of three clumps (E, NW and SW) emerged from this analysis, with the E one being the more massive component and coincident with the mass peak recovered by the weak lensing analysis. The authors suggested a complex merger history: a major merger along the E-W direction with a component along the line-of-sight gave origin to the E relic, while a minor merger along the N-S direction and parallel to the plane of the sky could be at the origin of the peculiar shape of the W relic. More recently, \citet{Golovich19a,Golovich19b} repeated a similar study confirming the results of \citet{Boschin10}.

\section{Data analysis}
\label{sec:analysis}

\subsection{Calibration and total intensity imaging}
\label{sec:calim}

A2345 has been observed with the JVLA in the L-band (1.008-2.032 GHz) B- and C-configurations. The bandwidth covers 1024 MHz, subdivided into 16 spectral windows of 64 MHz each (with 64 channels of 1 MHz frequency resolution). The observations have been performed with full polarization products. Central frequency, observing date and time of radio observations are listed in Tab.~\ref{tab:obs}.

\begin{table*}
    \centering
	\caption{Details of radio observations. Column 1: central observing frequency; Column 2: array configuration; Column 3: date of the observation; Column 4: total on-source observing time; Column 5:  Full Width Half Maximum (FWHM) of the major and minor axes of the restoring beam of the final total intensity image obtained with \texttt{robust}=0.5; Column 6: 1$\sigma$ rms noise of the total intensity image; Column 7: reference to the figures in this paper.}
	\label{tab:obs}
	\begin{tabular}{ccccccc} 
		\hline
		Freq.  & Array Conf. & Obs. Date & Obs. Time & Beam & $\sigma$ & Fig. \\
		(GHz)      &               &           &    (hr)       &       & (mJy/beam) & \\
		\hline
		1.5     &     B       &   2017 Nov. $\&$ Dec. &   4.0  &   3.3$\arcsec\times$4.8$\arcsec$ & 0.015 & Fig.~\ref{fig:opticsrc},~\ref{fig:opticE},~\ref{fig:opticW} \\
		1.5     &     C       &   2017 Jun.        &   1.5  &   11$\arcsec\times$18$\arcsec$ & 0.07 & Fig.~\ref{fig:X},~\ref{fig:opticE},~\ref{fig:opticW} \\
		\hline
	\end{tabular}
\end{table*}

We used the \texttt{CASA 5.6.2} package for the data reduction and total intensity imaging processes. Data were pre-processed by the VLA \texttt{CASA} calibration pipeline, that performs flagging and calibration procedures which are optimized for Stokes I continuum data. Then, we derived final delay, bandpass, gain/phase, leakage, and polarization angle calibrations. The sources used for the bandpass, absolute flux density, and polarization angle calibrations were 3C\,286 and 3C\,138. We used the \citet{Perley13} flux density scale for wide-band observations. We followed the NRAO polarimetry guide for polarization calibration\footnote{\url{https://science.nrao.edu/facilities/vla/docs/manuals/obsguide/modes/pol}}: a polynomial fit to the values of linear polarization fraction and polarization angle of 3C\,286 and 3C\,138 was used as frequency-dependent polarization model. The source J2131-1207 was used as phase calibrator for all the observations. The unpolarized sources J1407+2827 and 3C\,147 were used as instrumental leakage calibrators. The calibration tables were finally applied to the target. 

Radio frequency interference (RFI) was removed manually and using statistical flagging algorithms also from the cross-correlation products. Some spectral windows were entirely removed: those centered at 1.168, 1.232 and 1.552 GHz (i.e, spectral windows 2, 3, and 8) in B-configuration observations, and those centered at 1.232, 1.552 and 1.616 GHz (i.e., spectral windows 3, 8 and 9) in C-configuration. After RFI removal, we averaged the data sets in time down to 6 s and in frequency with channels of 4 MHz, in order to speed up the imaging and self-calibration processes. We computed new visibility weights according to their scatter.

Data have been imaged using the multi-scale multi-frequency de-convolution algorithm of the \texttt{CASA} task \texttt{tclean} \citep{Rau11} for wide-band synthesis-imaging. As a first step, we made a large image of the entire field ($\sim1^{\circ}\times1^{\circ}$). We used a three Taylor expansion (\texttt{nterms} = 3) in order to take into account both the source spectral index and the primary beam response at large distances from the pointing center. We also used the $w$-projection algorithm to correct for the wide-field non-coplanar baseline effect \citep{Cornwell08}. We set 128 and 64 $w$-projection planes for the B- and C-configuration data set, respectively. At this first stage, we used the uniform weighting scheme in order to minimize the synthesized beam side-lobes level, as well as to better image sources with high signal-to-noise ratios. The large images were then improved with several cycles of self-calibration to refine the antenna-based phase gain variations. During the last cycle amplitude gains were also computed and applied. The two observations performed in B-configuration were self-calibrated together. 

The second step was to subtract from the visibilities all the sources external to the field of interest ($\sim20\arcmin\times20\arcmin$). This was done, both, to reduce the noise generated by bright sources in the field and to speed up the subsequent imaging processes. Since the subtraction is not applied to cross-correlation products, polarized sources will be present outside the field of interest. This is not a problem since, both, the polarized flux density and the number of polarized sources are lower. After the subtraction, we used only two Taylor terms, we reduced the number of $w$-projection planes, and we set Briggs weighting scheme with the \texttt{robust} parameter set to 0.5. The latter choice was done to better image the extended emission. We performed a final cycle of phase and amplitude self-calibration. The final images were corrected for the primary beam attenuation using the \texttt{widebandpbcor} task in \texttt{CASA}. The residual calibration errors on the amplitude are estimated to be $\sim5 \ \%$. The restoring
beam and the local root mean square (rms) noise in the central region of the final images, $\sigma$, are listed in Tab.~\ref{tab:obs}.

\subsection{Polarization imaging}
\label{sec:polim}

To produce Stokes $I$, $Q$ and $U$ images for the polarization analysis, we used \texttt{WSCLEAN 2.8.1}\footnote{\url{https://sourceforge.net/p/wsclean/wiki/Home/}} \citep{Offringa14,Offringa17}.

We produced both full-band and sub-band images. The latter, with a frequency resolution of 16 MHz each, were used for the RM synthesis (see Sec.~\ref{sec:rmsynth}). The Stokes Q and U images were cleaned together using the \texttt{join-channels} and \texttt{join-polarizations} options. Full-band Stokes $I$ was used as a mask for the RM synthesis  and to compute the fractional polarization. We used the Briggs weighting scheme with \texttt{robust} = 0.5. The restoring beam was forced to be the same in the full-band image and in each frequency sub-band, matching the lowest resolution one (i.e., at 1.02 GHz). Each sub-band image was corrected for the primary beam calculated for the central frequency of the sub-band. The parameters describing the images used for the polarization analysis are listed in Tab.~\ref{tab:pol}.

\begin{table*}
    \centering
	\caption{Details of polarized intensity images. Column 1: array configuration; Column 2: central frequency of the first sub-band used for the RM synthesis; Column 3: central frequency of the last sub-band used for the RM synthesis; Column 4: width of the frequency sub-band used for the RM synthesis; Column 5: number of sub-bands used in the RM synthesis (excluding the flagged ones); Column 6: FWHM of the major and minor axes of the common restoring beam imposed to the sub-band and full-band images used for the polarization analysis (see Sec.~\ref{sec:polim}); Column 7: 1$\sigma$ rms noise of the full-band total intensity image; Column 8: average rms noise of polarized intensity images resulting from the $\widetilde{Q}(\phi)$ and $\widetilde{U}(\phi)$ spectra obtained with the RM synthesis. The average is computed over the image of the values of $\sigma_{QU}$ obtained for each unmasked pixel as $(\sigma_Q+\sigma_U)/2$ (see Sec~\ref{sec:rmsynth}); Column 9: reference to the figures in this paper.}
	\label{tab:pol}
	\begin{tabular}{ccccccccc} 
		\hline
		Array Conf. &  $\nu_i$ & $\nu_f$  & $\delta\nu$ & Sub-bands & Beam & $\sigma$ & <$\sigma_{QU}$> & Fig. \\
		       & (GHz) & (GHz) & (MHz) &   &   & (mJy/beam) & (mJy/beam/RMSF) &  \\
		\hline
		 B     & 1.015 & 2.023 & 16 & 48 &  8$\arcsec\times$8$\arcsec$ & 0.02 & 0.009 & Fig.~\ref{fig:relic_E},~\ref{fig:relic_W}  \\
		 C    & 1.015 & 2.023 &  16 & 49 & 30.5$\arcsec\times$30.5$\arcsec$ & 0.05 & 0.02 & Fig.~\ref{fig:X},~\ref{fig:relic_E},~\ref{fig:relic_W}\\
		\hline
	\end{tabular}
\end{table*}

Some frequency sub-band were discarded due to their higher noise with respect to average rms noise in the sub-bands: in Tab.~\ref{tab:pol} we list the number of sub-bands used for the RM synthesis for each configuration.

\subsection{RM synthesis}
\label{sec:rmsynth}

In this Section, we describe the procedure used to derive the RMs of the sources using the RM synthesis technique. We refer to \citet{Brentjens05} for a comprehensive introduction to this procedure. In the following, we will refer to the Faraday depth, $\phi$, to describe the Faraday space in which the RM synthesis is performed, but we will use the more common term RM to describe the actual value derived applying this technique. This is possible because we detected only Faraday-simple sources, which are not resolved in Faraday space.

We performed the RM synthesis on the $Q(\nu)$ and $U(\nu)$ sub-band images with \texttt{pyrmsynth}\footnote{\url{https://github.com/mrbell/pyrmsynth}}. We used equal weights for all the sub-bands and we imposed a spectral correction using an average spectral index $\alpha$ = 1. We obtained the reconstructed $\widetilde{Q}(\phi)$ and $\widetilde{U}(\phi)$ cubes in the Faraday space. Thus, in each pixel of the image, we obtained the reconstructed Faraday dispersion function, or Faraday spectrum, $\widetilde{F}(\phi)$. Faraday cubes were created between $\pm$1000 rad m$^{-2}$ and using bins of 2 rad m$^{-2}$.

\citet{Brentjens05} obtained approximated formulas to compute: the resolution in Faraday space, $\delta\phi$, the maximum observable Faraday depth, $|\phi_\text{max}|$, and the largest observable scale in Faraday space, $\Delta\phi_\text{max}$ (i.e., the depth and the $\phi$-scale at which sensitivity has dropped to $50 \ \%$). These parameters depend on the observational bandwidth and on the width of the sub-bands, which are listed in Tab.~\ref{tab:pol}. Therefore, in our case:

\begin{align}
    &\delta\phi \sim 45 \ \rm rad\,m^{-2} \ ,\\
    &|\phi_\text{max}| \sim 535 \ \rm rad\,m^{-2} \ ,\\
    &\Delta\phi_\text{max} \sim 143 \ \rm rad\,m^{-2} \ .
\end{align}

We masked the $Q(\nu)$ and $U(\nu)$ sub-band images using the full-band total intensity image: we thus run \texttt{pyrmsynth} only for those pixels above $3\sigma$ in total intensity. We also performed the RM clean down to the same threshold \citep[see][for the RM clean technique]{Heald09}. For each pixel we measured the noise of $\widetilde{Q}(\phi)$ and $\widetilde{U}(\phi)$ computing the rms, $\sigma_Q$ and $\sigma_U$, in the external ranges of the spectrum: at $|\phi| \ > \ 500 \ {\rm rad \ m}^{-2}$. This Faraday depth range is chosen to be outside of the sensitivity range of our observations (defined by |$\phi_\text{max}|$) and to avoid contamination from residual side-lobes of the sources. Since $\sigma_Q\sim\sigma_U$, we estimated the noise of each pixel of the polarization observations as $\sigma_{QU}=(\sigma_Q+\sigma_U)/2$ \citep[see also][]{Hales12}. By definition, $\sigma_{QU}$ is in units of Jy/beam/RMSF, where Rotation Measure Sampling Function (RMSF) represents the instrumental response in the Faraday space, similarly to the observing beam in the image domain. In Tab.~\ref{tab:pol}, we list the average value of $\sigma_{QU}$ for all the unmasked pixels in each observation.

We fitted pixel-by-pixel a parabola around the main peak of the Faraday spectrum. We thus obtained the RM (i.e., the Faraday depth at the peak, $\phi_{\rm peak}$) and polarized intensity ($|\widetilde{F}(\phi_{\rm peak})|$) images  from  the  coordinates  of  the  parabola  vertex in each pixel. For our analysis, we considered only pixels with a peak in the Faraday spectrum above a threshold of 6$\sigma_{QU}$. This corresponds to a Gaussian significance level of about 5$\sigma$ \citep[see][]{Hales12}.

The pixel-wise uncertainty on $\phi_{\rm peak}$ (and thus on the RM value in the single pixel) is derived following \citet{Brentjens05}, where:

\begin{equation}
    \sigma_{\phi}=\frac{\delta\phi}{2 P/\sigma_{QU}}~,
\label{eq:errphi}
\end{equation}
 
that is the FWHM of the RMSF divided by twice the signal-to-noise of the detection \citep[see also][]{Schnitzeler17}. We caution about this estimate since it is derived under the assumption of spectral index $\alpha$ = 0 and $\sigma_{Q}=\sigma_{U}$. In other cases, it can lead to over- or  under-estimates of the errors \citep{Schnitzeler17}.
 
We computed polarization intensity images using the peak of the Faraday dispersion function, and correcting for the Ricean bias as $P=\sqrt{|\widetilde{F}(\phi_{\rm peak})|^2-2.3\sigma_{QU}^2}$ \citep{George12}. We then obtained fractional polarization images dividing the $P$ images (with the 6$\sigma_{QU}$ threshold) by the full-band Stokes $I$ images (masked at the 3$\sigma$ level).

Our polarization images are not corrected for direction-dependent effects caused by the variation of the antenna primary beam pattern \footnote{This correction, named A-projection, has been very recently implemented in radio imaging software but has been validated only for a few usage modes. Currently, this still represents a limitation for wide-field polarization studies and deserves a huge effort from the radio-astronomical community.}. These effects can cause beam squint and off-axis flux leakage from the total intensity to the other Stokes parameters \citep{Bhatnagar13}. The strongest effect is visible in the Stokes I and V images. We estimated that in our images the V/I ratio increases with the distance from the pointing center, going from $1\%$, at a distance of $2\arcmin$, to $4\%$ at $12\arcmin$. This constrain the leakage to Stokes Q and U to be within 2$\%$ of the total intensity flux within the field of interest. This spurious contribution can be important for polarized sources with low fractional polarization. Therefore, we will not discuss the fractional polarization obtained for the sources observed in our field whenever it is below the 5$\%$ level. The instrumental leakage is centered on 0 rad m$^{-2}$ \citep[see][]{Jagannathan17}. Hence, we will consider the RMs as not affected by polarization leakage when |RM|>45 rad m$^{-2}$ (i.e., when the detected sources are at a distance of more than one RMSF from 0 rad m$^{-2}$). RMs below this threshold can differ from the true value by about $5\%$ \citep{Jagannathan17}. Hence, in this case, a $5\%$ uncertainty is added to the value computed with Eq.~\ref{eq:errphi}.

\subsection{X-ray data analysis}
\label{sec:Xray}

\begin{figure*}
	\includegraphics[width=\textwidth]{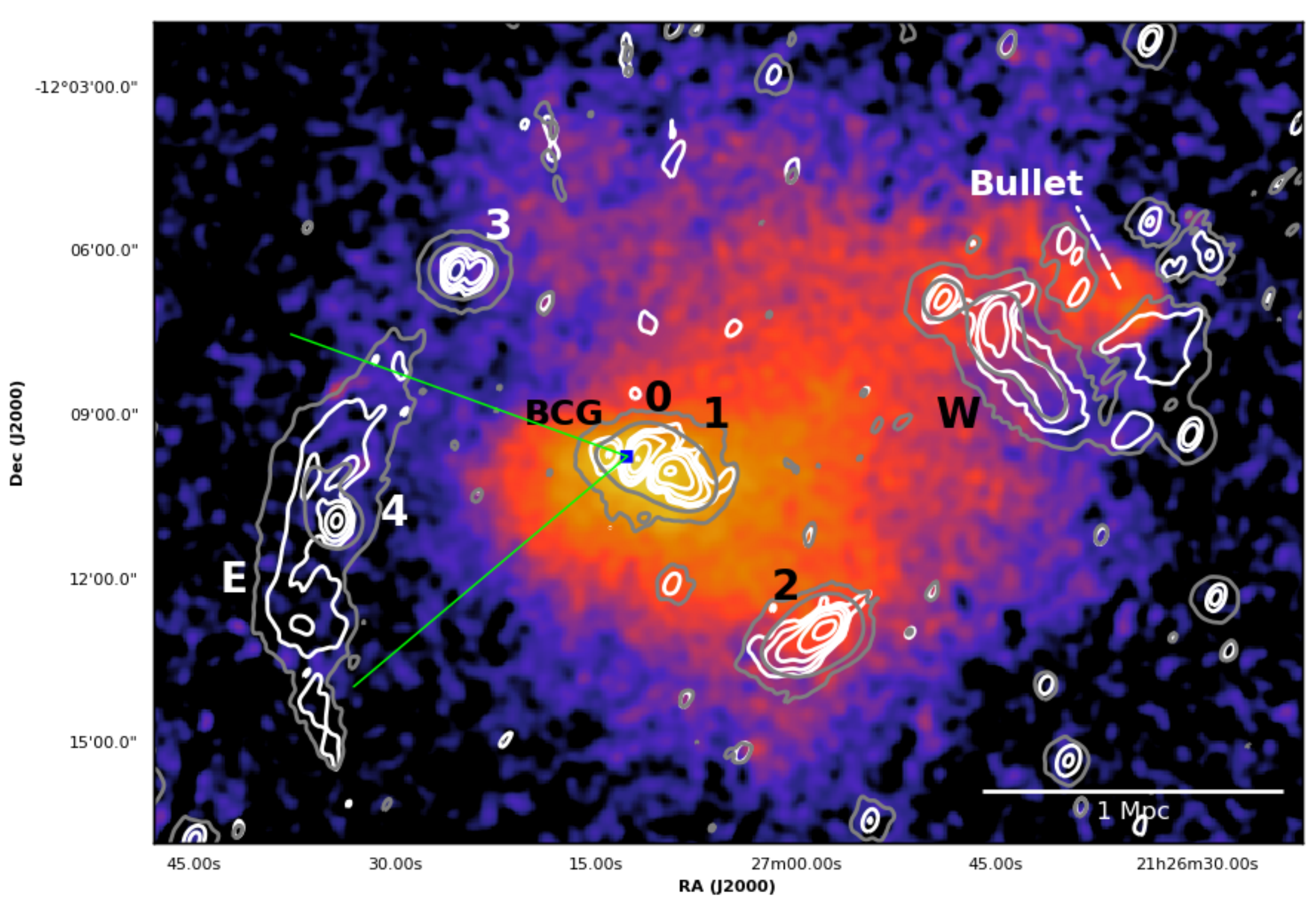}
    \caption{X-ray XMM-Newton point-source subtracted image of the cluster Abell 2345 (0.3-2 keV) with 1.5 GHz radio contours overlaid. Contours are from the C-configuration observation: white contours have a a restoring beam of 11$\arcsec\times$18$\arcsec$, while gray contours show the same data set with a restoring beam of 30.5$\arcsec\times$30.5$\arcsec$. White contours start at 3$\sigma$ and are spaced by a factor of four, where $\sigma$ is the value listed in Tab.~\ref{tab:obs}. Gray contours show only the 3$\sigma$ and 48$\sigma$ levels and $\sigma$ is the value listed in Tab.~\ref{tab:pol}. The five radio galaxies detected in polarization are marked with numbers, while the eastern and western relics are marked with the letters ‘‘E’’ and ‘‘W’’, respectively. The blue square marks the position of the X-ray surface brightness peak, at the J2000 coordinates: 21$^{\rm h}$27$^{\rm m}$12$^{\rm s}$.6, -12$^{\circ}$09$\arcmin$46$\arcsec$. Green lines show the boundaries of the sector used to extract the surface brightness profile and to model the thermal electron density distribution.}
    \label{fig:X}
\end{figure*}

A2345 was observed by XMM-Newton in April 2010 during rev. 1900  (ObsID: 0604740101) with a total exposure time of ~ 93 ks. The observation was performed in full frame mode for the MOS cameras and extended full frame mode for the pn detector, all using the thin filter.

Observation data files (ODFs) were downloaded from the XMM-Newton archive and processed with the \texttt{XMMSAS 16.0.0} software for data reduction \citep{Gabriel04}. We used the tasks \texttt{emchain} and \texttt{epchain} to generate calibrated event files from raw data. We excluded all the events with \texttt{PATTERN$>$4} for pn data and with \texttt{PATTERN$>$12} for MOS data. In addition, bright pixels and hot columns were removed in a conservative way by applying the expression \texttt{FLAG==0}. We discarded the data corresponding to the periods of high background induced by solar flares using the two-stage filtering process extensively described in \citet{Lovisari11}. The remaining exposure times after cleaning are 47.5 ks for MOS1, 51.5 ks for MOS2,  and 25.5 ks for pn. Point-like sources were detected using the task \texttt{edetect-chain} and excluded from the event files. The background event files were cleaned by applying the same \texttt{PATTERN} selection, flare rejection criteria, and point-source removal used for the observation events. The resulting image is shown in Fig.~\ref{fig:X}.

The X-ray morphology of A2345 is strongly disturbed and an average surface brightness profile would result in a poor description of the thermal environment of each source. The deviation from spherical symmetry is stronger in the north-western side of the cluster and far away from the BCG, confirming weak lensing studies \citep[e.g.,][]{Dahle02,Okabe10}. We used the background-subtracted and exposure-corrected images in the 0.3-2 keV energy band to extract the surface-brightness profiles in a sector centered on the X-ray peak, and encompassing the radio sources of interest. In particular, we are interested in the sector containing the E relic in order to study the magnetic field profile up to the relic region. This sector is also the less disturbed one (see Fig.~\ref{fig:X}).

A double $\beta$-model was used for fitting:
\begin{equation}
    S_{X}(r) = S_{X,1} \bigg[ 1+ \bigg( \frac{r}{r_{c,1}} \bigg)^2 \bigg]^{-3\beta_1+0.5} + S_{X,2} \bigg[ 1+ \bigg( \frac{r}{r_{c,2}} \bigg)^2 \bigg]^{-3\beta_2+0.5} \ ,
\label{eq:2betasb}
\end{equation}

where the central surface brightness, $S_{X,i}$, the core radius, $r_{c,i}$, and the $\beta_i$ parameter of each component were left free to vary. A single $\beta$-model would provide a poor description of the surface brightness profile in the considered sector.

Under the assumption of spherical symmetry, the electron density profile in the sector can be obtained combining spectral (i.e., using the normalization of an APEC model obtained by fitting a spectrum extracted in the sector) and imaging analysis (i.e., the best-fits values of the double $\beta$-model), as described in \citet{Lovisari15} \citep[see also][]{Hudson10}. For a double $\beta$-model the thermal electron density profile is: 
\begin{equation}
    n_e(r) = \bigg\{ n_{e,1}^2 \bigg[ 1+ \bigg( \frac{r}{r_{c,1}} \bigg)^2 \bigg]^{-3\beta_1} + n_{e,2}^2 \bigg[ 1+ \bigg( \frac{r}{r_{c,2}} \bigg)^2 \bigg]^{-3\beta_2} \bigg\}^{0.5} \ ,
\label{eq:2betadensity}
\end{equation}
where $n_{e,1}$ and $n_{e,2}$ are the central densities of the two components. Indeed, due to the complex structure of A2345 the assumption of spherical symmetry is a source of uncertainty in our modeling. However, we note that using a narrow sector for the calculation of the profile helps to mitigate this effect.

\section{Results}
\label{sec:reluts}

\begin{figure}
	\includegraphics[width=\columnwidth]{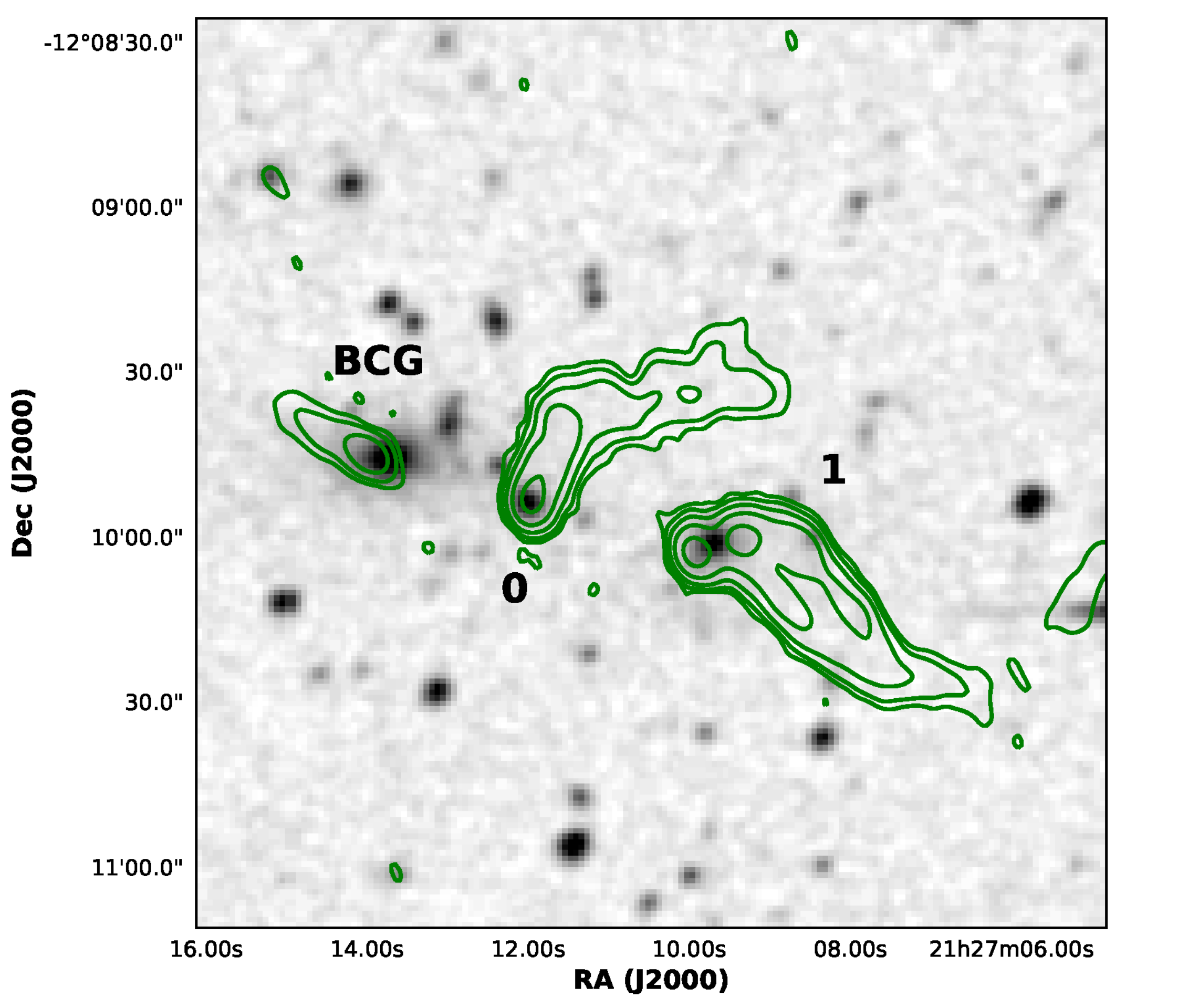}
    \caption{Optical DSS2 image of the central region of A2345 with 1.5 GHz radio contours overlaid. Contours are from the B-configuration observation with a restoring beam of 3.3$\arcsec\times$4.8$\arcsec$. Contours start at $3\sigma$ and are spaced by a factor of four. The value of $\sigma$ is listed in Tab.~\ref{tab:obs}. The three central sources are marked with the same labels of Fig.~\ref{fig:X}.}
    \label{fig:opticsrc}
\end{figure}

\begin{figure}
	\includegraphics[width=\columnwidth]{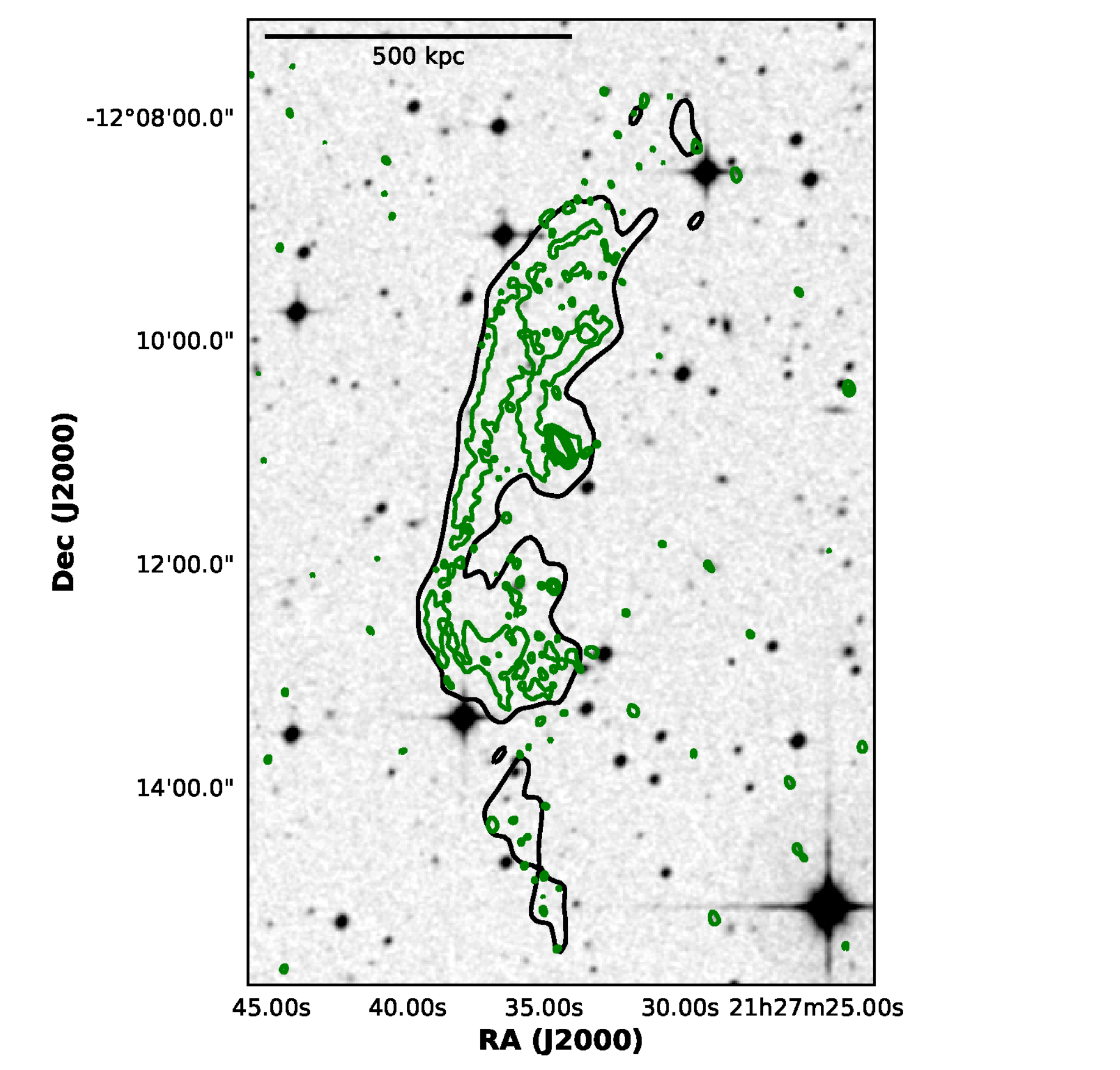}
    \caption{Optical DSS2 image of the E relic with 1.5 GHz radio contours overlaid. Green contours are from the B-configuration observation. They have a restoring beam of 3.3$\arcsec\times$4.8$\arcsec$, start at $3\sigma$ and are spaced by a factor of four. The value of $\sigma$ is listed in Tab.~\ref{tab:obs}. The black contour is the $3\sigma$ level of same C-configuration observation shown with white contours in Fig~\ref{fig:X}.}
    \label{fig:opticE}
\end{figure}

\begin{figure}
	\includegraphics[width=\columnwidth]{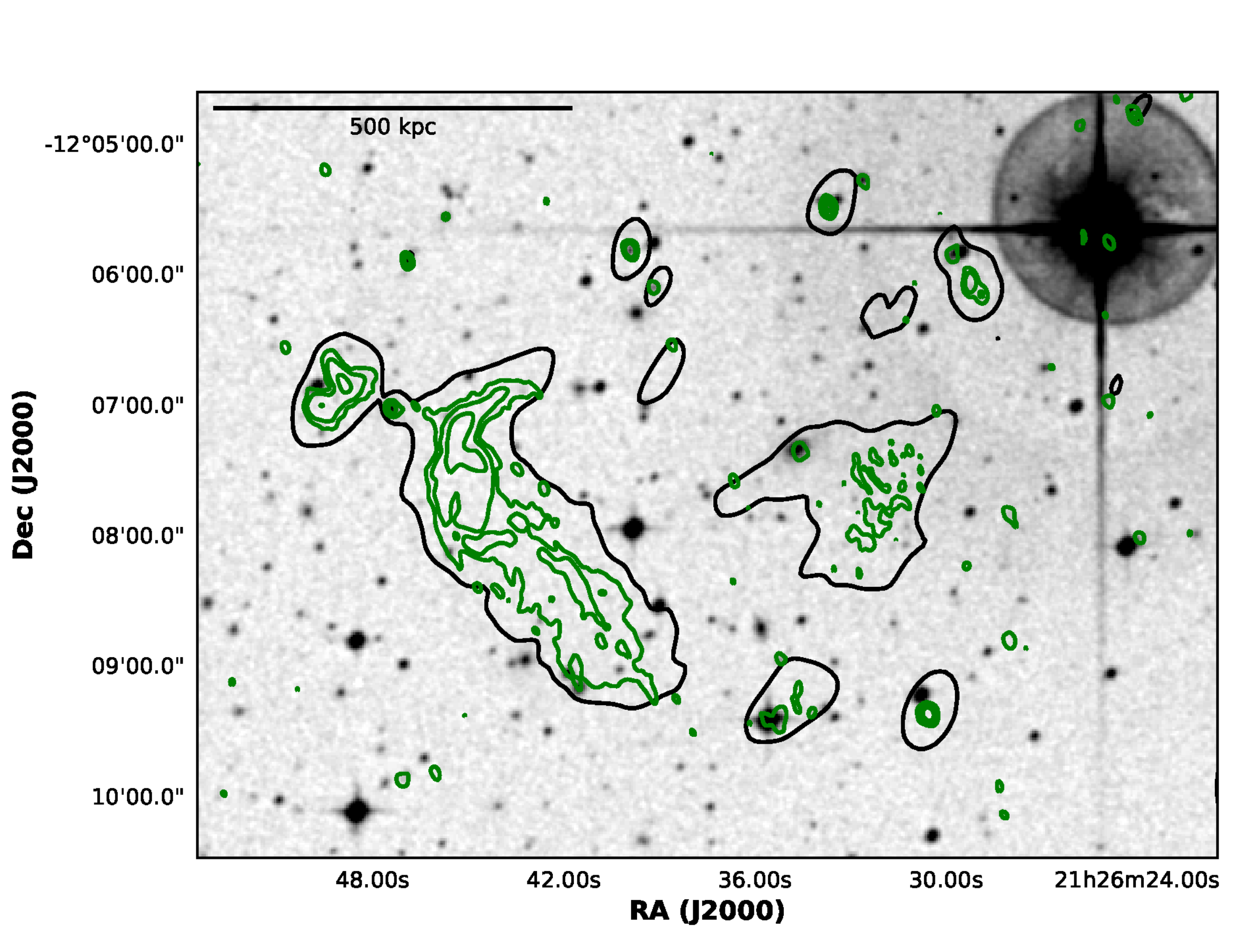}
    \caption{Same as Fig:~\ref{fig:opticE} for the W relic.}
    \label{fig:opticW}
\end{figure}

The XMM-Newton image of A2345 is shown in Fig.~\ref{fig:X}, overlaid with C-configuration total intensity contours. The central core is elongated in the NE-SW direction while a northern bullet-like component has a peak in the NW and shows an elongated tail toward the eastern direction. 

Three radio sources lie nearby the X-ray surface brightness peak (at the position listed in Tab.~\ref{tab:info}) and are all resolved in the B-configuration observation (see Fig.~\ref{fig:opticsrc}). The eastern one is the BCG identified by \citet{Boschin10}, at redshift $z$=0.181, the other two (marked as source 0 and source 1) are tailed radio sources. The source 0 is a narrow angle tail radio galaxy \citep[NAT, e.g.][]{Miley80} while the presence of two warm-spots and of two distinguishable tails suggest the wide angle tail (WAT) classification for the source 1 \citep[e.g.,][]{Missaglia19}. Another tailed radio galaxy (marked as source 2 in Fig.~\ref{fig:X}) lies 5.2$\arcmin$ (i.e., $\sim$950 kpc) away from the BCG to the SW direction. These classes of sources are commonly found at the center of merging galaxy clusters, where the dynamic pressure resulting from their motion through the surrounding ICM swept back their jets \citep{Sakelliou00}. The tails of these sources point toward different directions, suggesting that they are on different radial orbits around the main potential well (see also Fig.~\ref{fig:opticsrc}).

The E radio relic is elongated along the N-S direction with a largest linear size of 1.41 Mpc (7$\arcmin$.7). It lies at a distance of $\sim$1 Mpc from the BCG, in a region of low X-ray surface brightness. The high-resolution image of the relic is shown in Fig.~\ref{fig:opticE}, overlaid on the optical image from the Second Digitized Sky Survey \citep[DSS2,][]{McLean00}. This image reveals the internal filamentary structure of the relic, representing a great improvement with respect to the observations performed by \citet{Bonafede09a}. In particular, a bright internal arc-like structure, with a linear size of 250 kpc and a transverse size of $\sim$25 kpc, is detached from the main large scale arc. A double lobed source is detected in the down-stream region of the relic, marked as source 4 in Fig.~\ref{fig:X}. 

The W radio relic has a peculiar morphology, as already noticed by \citet{Bonafede09a}. It lies at a distance of $\sim$1.3 Mpc from the central BCG. In our high-resolution image (see Fig.~\ref{fig:opticW}) it shows a main structure elongated for 455 kpc (2$\arcmin$.5) in the NE-SW direction. At the northern edge this structure is connected with an arc-like filament elongated in the perpendicular direction. It is difficult to judge whether this arc is purely diffuse emission, or a radio galaxy with a faint counterpart visible in the DSS2 image. There are diffuse patches of radio emission also toward the outskirts of the cluster. The faint outer emission is visible also in the upper-right corner of Fig.~\ref{fig:X} and it surrounds the NW X-ray peak. Although at low resolution this emission gets blended with a number of point-like sources, we checked that the flux measured from the low resolution image has a higher flux density (38.5 mJy) with respect to the sum of the flux densities measured from single point-like sources detected in the high resolution image (3.9 mJy). In particular, the largest patch of emission coincides with the position of the bullet-like X-ray structure, likely generated by a sub-cluster motion toward W. 

\subsection{Polarized radio galaxies}
\label{sec:polgal}

The polarization analysis of the B-configuration observation allowed the detection of five radio galaxies within the field (marked with numbers from 0 to 4 in Fig.~\ref{fig:X}). Sources 0, 1 and 2 are confirmed cluster members \citep{Boschin10}, while 3 and 4 are likely background radio sources. For each of them, we computed a pixel-wise average RM, <RM>, using only pixels detected with a signal-to-noise ratio higher than 6, as specified in Sec.~\ref{sec:rmsynth}. We computed the RM dispersion, $\sigma_\text{RM}$, for each source as $\sqrt{\sigma_{\rm RM,obs}^2-{\rm med}(\sigma_{\phi})^2}$, where $\sigma_{\rm RM,obs}$ is the observed standard deviation of the pixels and ${\rm med}(\sigma_{\phi})$ is the median error of the RM estimate at each pixel as in Eq.~\ref{eq:errphi}. The value of ${\rm med}(\sigma_{\phi})$ is $\sim1-3$ rad m$^{-2}$ for all the sources. The estimates of <RM> and $\sigma_{RM}$ are listed in Tab.~\ref{tab:sources}. Moreover, we listed the median RM value of each source and the median absolute deviation (MAD), which are good estimators in the case of low statistics and presence of outliers due to low signal-to-noise in the sampled regions. We also listed in Tab.~\ref{tab:sources} the number of resolution beams, $n_\text{beam}$, sampled by each source in polarization with a 6$\sigma_{QU}$ detection threshold.

\begin{table*}
    \centering
\caption{Polarization properties of the sources detected in polarization in the B-configuration observation. Column 1: identification number of the source as shown in Fig.~\ref{fig:X}; Column 2,3: J200 celestial coordinates of the source measured at the position of the brightest polarized pixel; Column 4: redshift of the source from \citet{Boschin10}; Column 5: average RM of the source; Column 6: standard deviation of the RM distribution after the subtraction of ${\rm med}(\sigma_{\phi}$); Column 7: median RM of the source; Column 8: median absolute deviation of the RM distribution; Column 9: median of the uncertainty on $\phi_{\rm peak}$, $\sigma_{\phi}$, for the considered pixels; Column 10: number of resolution beams covered by the pixels detected above a 6$\sigma_{QU}$ detection threshold in the B-configuration observation, rounded to a whole number; Column 11: distance of the source from the X-ray surface brightness peak. All the statistical quantities are computed using only pixels with signal-to-noise ratio higher than 6 in polarization.}
\begin{tabular}{|c|c|c|c|c|c|c|c|c|c|c|}
\hline
  \multicolumn{1}{|c|}{Source} &
  \multicolumn{1}{c|}{R.A.} &
  \multicolumn{1}{c|}{Dec.} &
  \multicolumn{1}{c|}{$z$} &
  \multicolumn{1}{c|}{<RM>} &
  \multicolumn{1}{c|}{$\sigma_\text{RM}$} &
  \multicolumn{1}{c|}{med(RM)} &
  \multicolumn{1}{c|}{MAD(RM)} &
  \multicolumn{1}{c|}{${\rm med}(\sigma_{\phi}$)} &
  \multicolumn{1}{c|}{$n_\text{beam}$} &
  \multicolumn{1}{c|}{Distance} \\
  \multicolumn{1}{|c|}{} &
  \multicolumn{1}{c|}{(deg)} &
  \multicolumn{1}{c|}{(deg)} &
  \multicolumn{1}{c|}{} &
  \multicolumn{1}{c|}{(rad m$^{-2}$)} &
  \multicolumn{1}{c|}{(rad m$^{-2}$)} &
  \multicolumn{1}{c|}{(rad m$^{-2}$)} &
  \multicolumn{1}{c|}{(rad m$^{-2}$)} &
  \multicolumn{1}{c|}{(rad m$^{-2}$)} &
  \multicolumn{1}{c|}{} &
  \multicolumn{1}{c|}{(kpc)} \\
\hline
  0 & 321.800 & -12.165 & 0.180 & 128 & 173 & 107 & 179 & 3 & 3 & 37\\
  1 & 321.789 & -12.167 & 0.179 & -80 & 131 & -62 & 57 & 3 & 3 & 154\\
  2 & 321.738 & -12.214 & 0.176 & 25 & 61 & 38 & 31 & 1 & 20 & 911\\
  3 & 321.850 & -12.107 & - & -11 & 35 & -28 & 9 & 2 & 2 & 816\\
  4 & 321.893 & -12.181 & - & 2 & 5 & 3 & 1 & 1 & 5 & 1015\\
\hline\end{tabular}
\label{tab:sources}
\end{table*}

\subsection{Polarization properties of the relics}
\label{sec:polrel}

\begin{figure*}
	\includegraphics[width=1\linewidth]{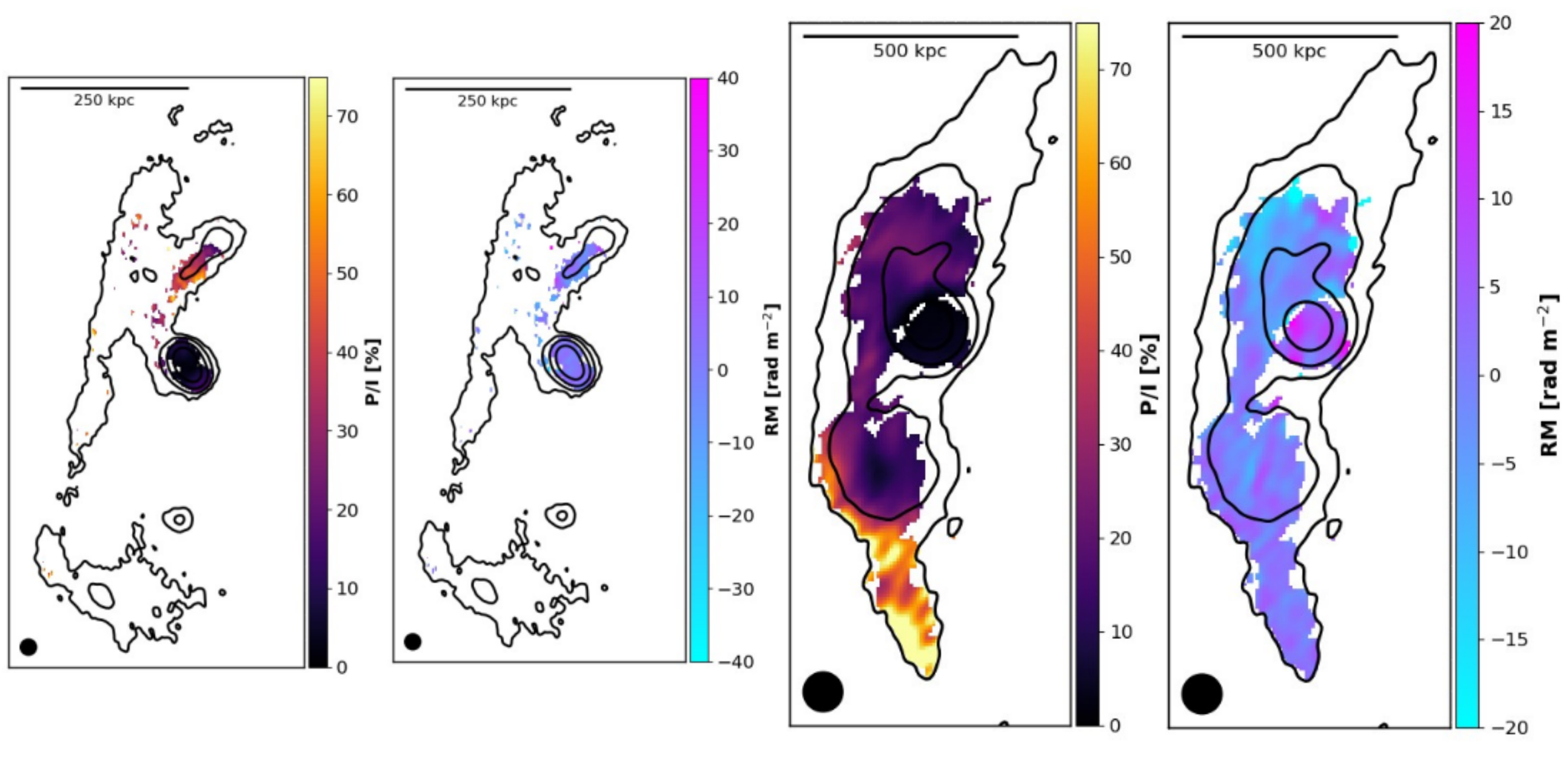}
\caption{Fractional polarization and RM images of the E radio relic in the B- and C-configuration observations. The $6\sigma_{QU}$ detection threshold was imposed in polarization and only pixels above this threshold are shown. Black contours show the total intensity image used to compute the fractional polarization, start from 3 times the rms noise and are spaced by a factor of four (more details on the images in Tab.~\ref{tab:pol}).}
    \label{fig:relic_E}
\end{figure*}

\begin{figure*}
    \includegraphics[width=1\linewidth]{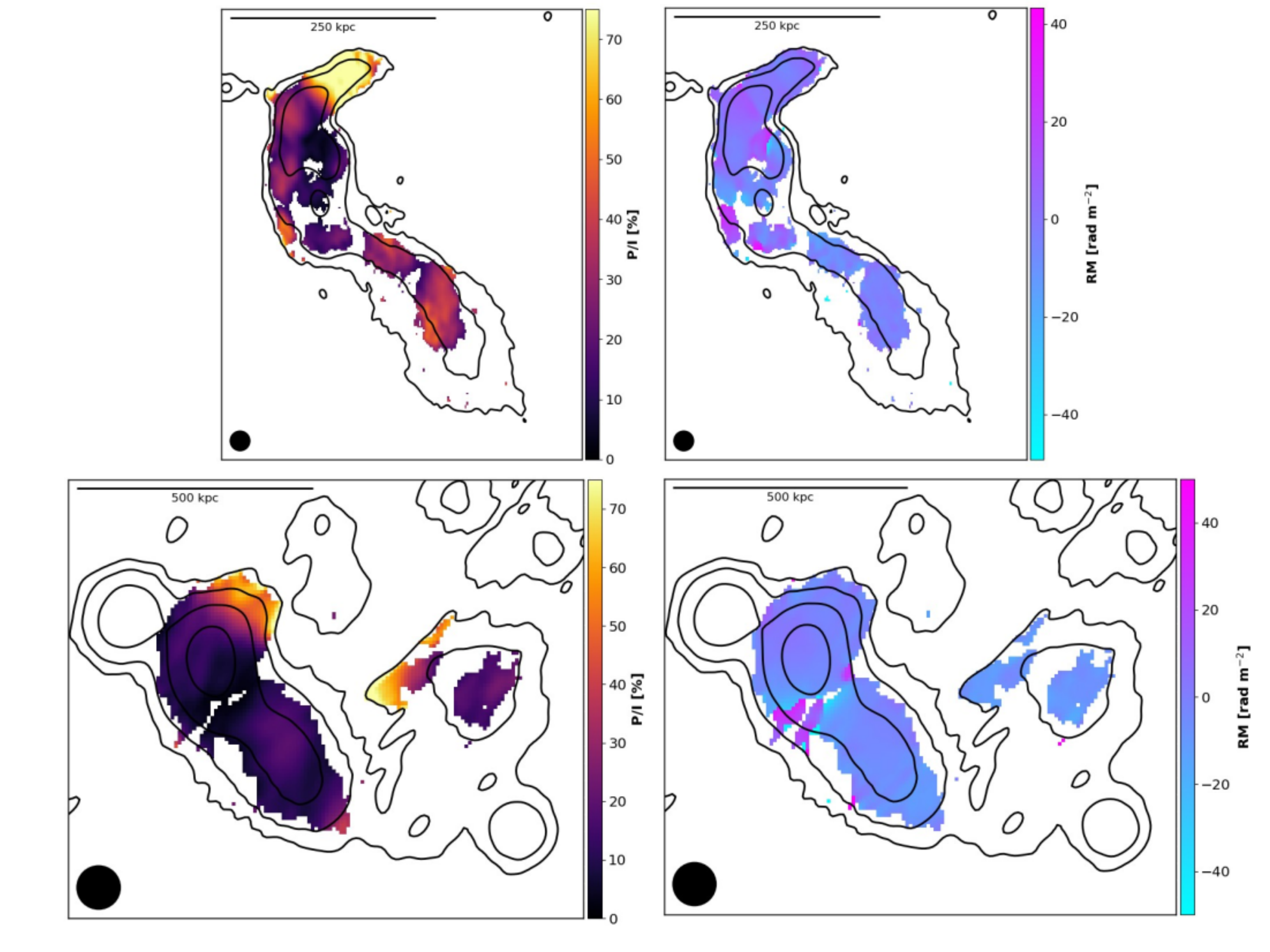}
    \caption{Same as Fig.~\ref{fig:relic_E} for the W relic.}
\label{fig:relic_W}
\end{figure*}

The two relics are detected in polarization at both high (i.e., B-configuration) and low (i.e., C-configuration) resolution. The extended emission is better recovered with the C-configuration observations, in particular for the E relic. The RM and fractional polarization images of the relics are shown in Fig.~\ref{fig:relic_E} and  \ref{fig:relic_W}. The same information obtained for polarized sources are listed in Tab.~\ref{tab:relics} for the relics, at both resolutions.

For the relics, that have negligible polarized flux leakage because of their high polarization, we also computed their average fractional polarization. We integrated the total intensity ($I$) and polarized ($P$) flux densities over the area covered by pixels detected above the 6$\sigma_{QU}$ detection threshold in polarization and we computed $P/I$. The uncertainty on the fractional polarization was computed as $P/I \cdot \sqrt{(\sigma_P/P)^2+(\sigma_I/I)^2}$ with the error on the flux densities estimated as

\begin{equation}
    \sigma_\text{flux}=\sqrt{(\delta f \cdot \text{flux})^2+(\text{noise} \cdot \sqrt{n_\text{beam}})^2} \, ,
\end{equation}

where flux = $I$, $P$ and noise = $\sigma$, <$\sigma_{QU}$> for total intensity and polarized flux densities, respectively. $\delta f$ is the residual calibration error on the flux ($5 \ \%$ for JVLA data) and $n_\text{beam}$ is the number of beam in the sampled region.

The E relic shows few polarized regions above the 6$\sigma_{QU}$ threshold in the B-configuration image (Fig.~\ref{fig:relic_E}, left panels). Most of the detected pixels coincide with the internal thin arc of this relic. The fractional polarization reaches the $65\%$ level here, and the average value is $34\pm3 \ \%$. The Faraday depth ranges between -28 and 45 rad m$^{-2}$ with a median RM of $-2 \ {\rm rad \ m^{-2}}$ and MAD(RM) = $5 \ {\rm rad \ m^{-2}}$. In the low-resolution C-configuration observation the extended emission of this relic is better sampled (see Fig.~\ref{fig:relic_E}, right panels). The polarized emission covers the entire relic, except for the northern region. The average fractional polarization is lower than at higher resolution (i.e., $18\pm1 \ \%$) but it reaches the $70\%$ in the southern part. We notice a decrease of the fractional polarization where the total intensity high-resolution image shows more substructures. In particular, in the region surrounding the internal thin arc, the decrease of polarized emission coincides with strong variation of the Faraday depth. In the southern region of the relic, that shows higher fractional polarization, RM variations are smoother than in the northern part. Therefore the depolarization is likely to be caused by substructures in the shock surface or the magnetic field within the beam. The median RM of the E relic at low-resolution is consistent with zero.

The polarized emission of the W relic is patchy and reaches the $75\%$ level in the northern region in the B-configuration observation (Fig.~\ref{fig:relic_W}, top panels). This emission could be associated to the lobe of a radio galaxy, but such a high level of fractional polarization is suggestive of a very ordered magnetic field which is expected in radio relics. The RM distribution in this region is smooth, while it is less homogeneous in the central part, indicating the presence of more substructures that causes depolarization. The average fractional polarization is $24\pm2 \ \%$. The degree of polarization decreases at low resolution but still reaches the $70\%$ in the northern part, with an average value of $12.6\pm0.9 \ \%$. It is interesting that, at low resolution, the patch of emission in front of the bullet-like X-ray structure appears to be polarized with a maximum polarization fraction of $73\%$ (Fig.~\ref{fig:relic_W}, bottom panels). This may suggest the presence of a shock which is ordering the magnetic field lines and thus increasing the fractional polarization in this region. An X-ray surface brightness jump is also visible at this position in Fig.~\ref{fig:X}. The median RM is $-2 \ {\rm rad \ m^{-2}}$ in the B-configuration observation, while it is $-5 \ {\rm rad \ m^{-2}}$ in the C-configuration. The MAD(RM) is $4 \ {\rm rad \ m^{-2}}$ in the B-configuration observation and $3 \ {\rm rad \ m^{-2}}$ in C-configuration.

In general, the obtained fractional polarization is consistent with the work of \citet{Bonafede09a}. The resolution achieved in this previous work was in fact intermediate between the ones of our high and low resolution images and the fractional polarization obtained by the authors has an intermediate value. This is expected since a larger amount of depolarization is generated when the polarized emission is mixed inside a larger observing beam. The differences in the average fractional polarization and in the morphology of the detections observed between B- and C-configuration are consistent with beam depolarization and with the different sensitivity obtained with the change in resolution.

\begin{figure}
    \includegraphics[width=1\columnwidth]{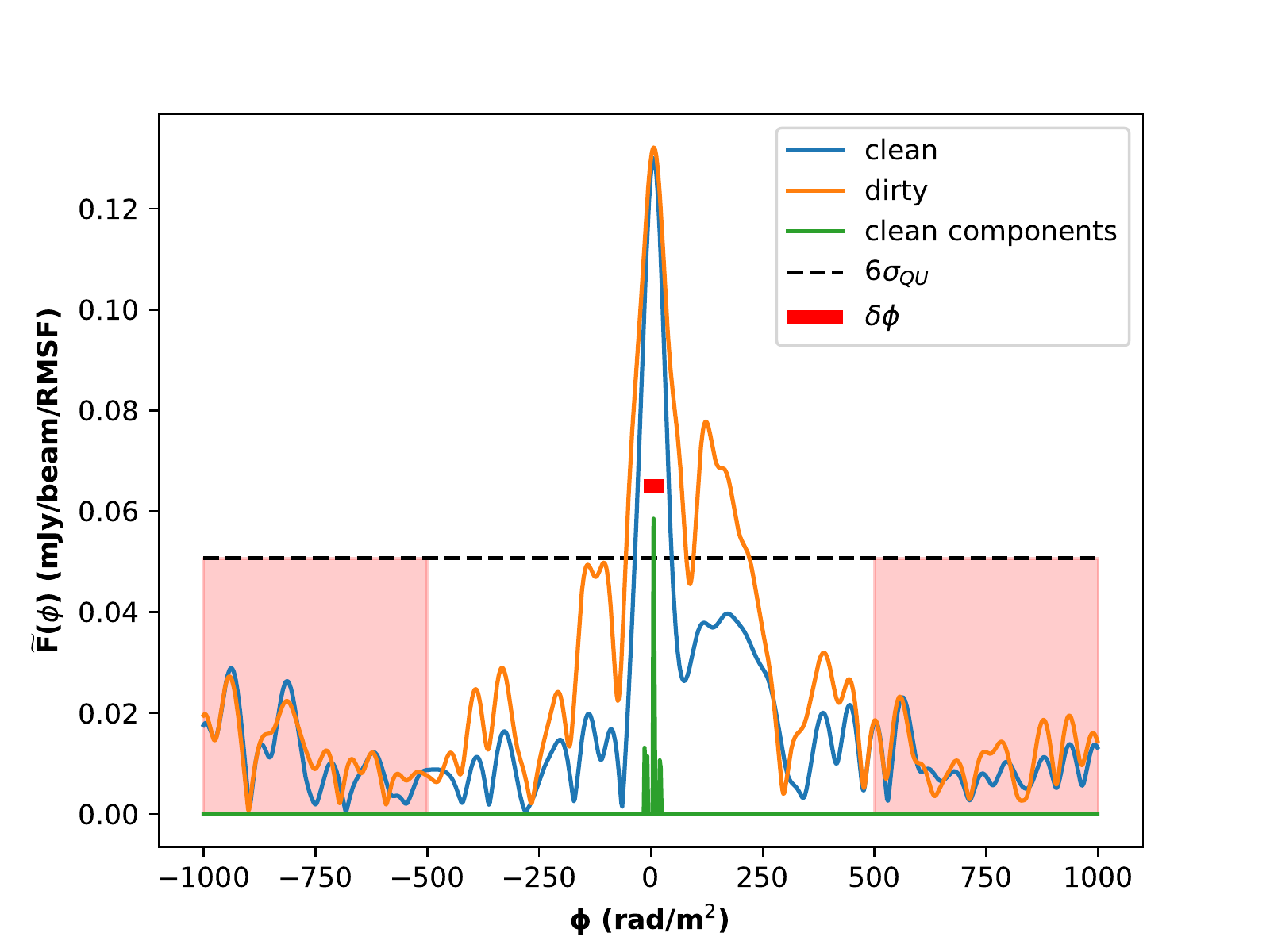}
    \includegraphics[width=1\columnwidth]{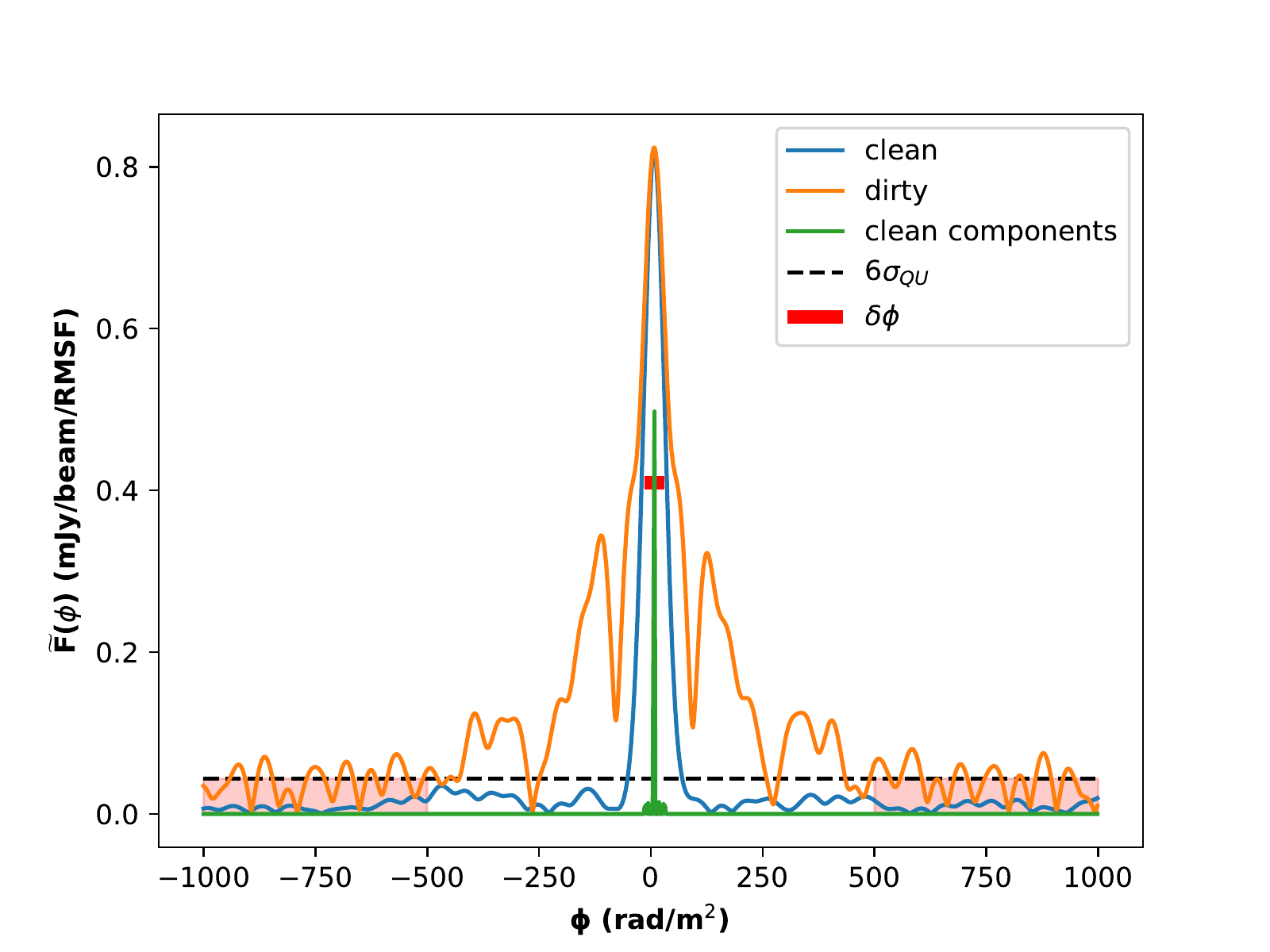}
    \caption{Faraday spectra of the brightest polarized pixels of the E relic (top panel) and of the W relic (bottom panel). The orange line is the dirty spectrum, the blue line is the spectrum after the RM clean, and the green lines show the cleaned components. For reference, the $6\sigma_{QU}$ detection threshold is plotted with a black dotted line and the red shadowed regions show the range of the spectrum where $\sigma_{QU}$ was computed. The width corresponding to the resolution in Faraday space is plotted at the half-maximum of the spectrum to show that the emission is Faraday-simple.}
\label{fig:pixels}
\end{figure}

The Faraday spectra detected from the two relics are Faraday-simple, meaning that they show a single peak with a FWHM coincident with the resolution of our observation in Faraday space (i.e., $\delta\phi\sim45 \ {\rm rad \ m^{-2}}$). As an example, the Faraday spectra of the brightest polarized pixels in the E and W relics are shown in Fig.~\ref{fig:pixels}. Several layers of radio-emitting plasma are expected to be present in radio relics and they may be unveiled by the RM synthesis  \citep[see, e.g.,][]{Stuardi19}. In this case, it is possible that the emitting layers of the relics are not resolved and that we detect only an external Faraday rotating screen.

\begin{table*}
    \centering
\caption{Polarization properties of the relics. Column 1: Array configuration;  Column 2: name of the relic as identified in Fig.~\ref{fig:X}; 
Column 3: average RM of the source; Column 4: standard deviation of the RM distribution after the subtraction of ${\rm med}(\sigma_{\phi}$); Column 5: median RM of the source; Column 6: median absolute deviation of the RM distribution; Column 7: median of the uncertainty on $\phi_{\rm peak}$, $\sigma_{\phi}$, for the considered pixels; Column 8: average fractional polarization with statistical uncertainties quoted in the $\pm1\sigma$ range; Column 9: number of resolution beams covered by the pixels detected above a 6$\sigma_{QU}$ detection threshold, rounded to a whole number; Column 10: distance of the relic from the X-ray surface brightness peak. All the statistical quantities are computed using only pixels with signal-to-noise ratio higher than 6 in polarization.}
\begin{tabular}{|c|c|c|c|c|c|c|c|c|c|}
\hline
  \multicolumn{1}{|c|}{Array Conf.} &
  \multicolumn{1}{|c|}{Relic} &
  \multicolumn{1}{c|}{<RM>} &
  \multicolumn{1}{c|}{$\sigma_\text{RM}$} &
  \multicolumn{1}{c|}{med(RM)} &
  \multicolumn{1}{c|}{MAD(RM)} &
  \multicolumn{1}{c|}{${\rm med}(\sigma_{\phi})$} &
  \multicolumn{1}{|c|}{$P/I$} &
  \multicolumn{1}{c|}{$n_\text{beam}$} &
  \multicolumn{1}{c|}{Distance} \\
  \multicolumn{1}{|c|}{} &
  \multicolumn{1}{|c|}{} &
  \multicolumn{1}{c|}{(rad m$^{-2}$)} &
  \multicolumn{1}{c|}{(rad m$^{-2}$)} &
  \multicolumn{1}{c|}{(rad m$^{-2}$)} &
  \multicolumn{1}{c|}{(rad m$^{-2}$)} &
  \multicolumn{1}{c|}{(rad m$^{-2}$)} &
  \multicolumn{1}{c|}{(\%)} &
  \multicolumn{1}{c|}{} &
  \multicolumn{1}{c|}{(Mpc)} \\
\hline
  B & E & 1 & 41 & -2 & 5 & 3 & 34$\pm$3 & 6 & 1.0\\
  B & W & -1 & 9 & -2 & 4 & 2 & 24$\pm$2 & 37 & 1.3 \\
  C & E & -1 & 6 & -0.2 & 2 & 2 & 18$\pm$1 & 18 & 1.0\\
  C & W & -4 & 13 & -5 & 3 & 2 & 12.6$\pm$0.9 & 14 & 1.3 \\
\hline\end{tabular}
\label{tab:relics}
\end{table*}

\subsection{The Galactic contribution}
\label{sec:Gal}

The mean RM value of the Galactic foreground in the region of the cluster is consistent with zero \citep[i.e., -0.2$\pm$5.2 rad m$^{-2}$,][]{Huts20}. Therefore, the Galactic contribution will not be subtracted out from our measurements. Nevertheless, when studying the RM distribution of sources in the cluster, the Galactic RM variance generated by the turbulence of the inter-stellar medium on the angular scales of the cluster should be considered. The Galactic RM variance has a strong dependence on angular separation and Galactic latitude \citep{Simonetti84,Simonetti92}. The largest angular distance between two polarized sources in our sample (i.e., the distance between the relics) is $\sim13\arcmin$. Although sub-degree angular scales are not well sampled by actual studies, the amount of Galactic RM variance on $\sim10\arcmin-15\arcmin$ scales is of the order of $\sim$10 rad m$^{-2}$, depending on the Galactic latitude \citep{Stil11}. For example, using the analytical formula derived by \citet{Anderson15}, Eq. 20, we can estimate the Galactic RM variance to be $\sim7$ rad m$^{-2}$ at $13\arcmin$. The standard deviation computed between the <RM> of the sources in A2345 (considering also the relics) is instead $\sim57$ rad m$^{-2}$, and thus this value cannot be entirely attributed to the Milky Way.

The median RM computed for the relics and from source 4 in the B-configuration observation is consistent with the Galactic mean RM. In the C-configuration, the med(RM) of the W relic is larger but still consistent with the Galactic one, due to the large uncertainty on the latter. Local enhancement of the RM within the regions of the relics are likely due to the local ICM and can be regarded as a small fluctuation around the mean, which is instead determined by the Faraday rotation within our Galaxy. The RM dispersion, or the MAD(RM), computed on the scales of the sources (i.e., angular scales below 2$\arcmin$.5, which is the angular extent of the E relic) are thus more indicative of the cluster magnetic field.

\subsection{RM profiles}
\label{sec:RMprof}

\begin{figure*}
    \includegraphics[width=1\linewidth]{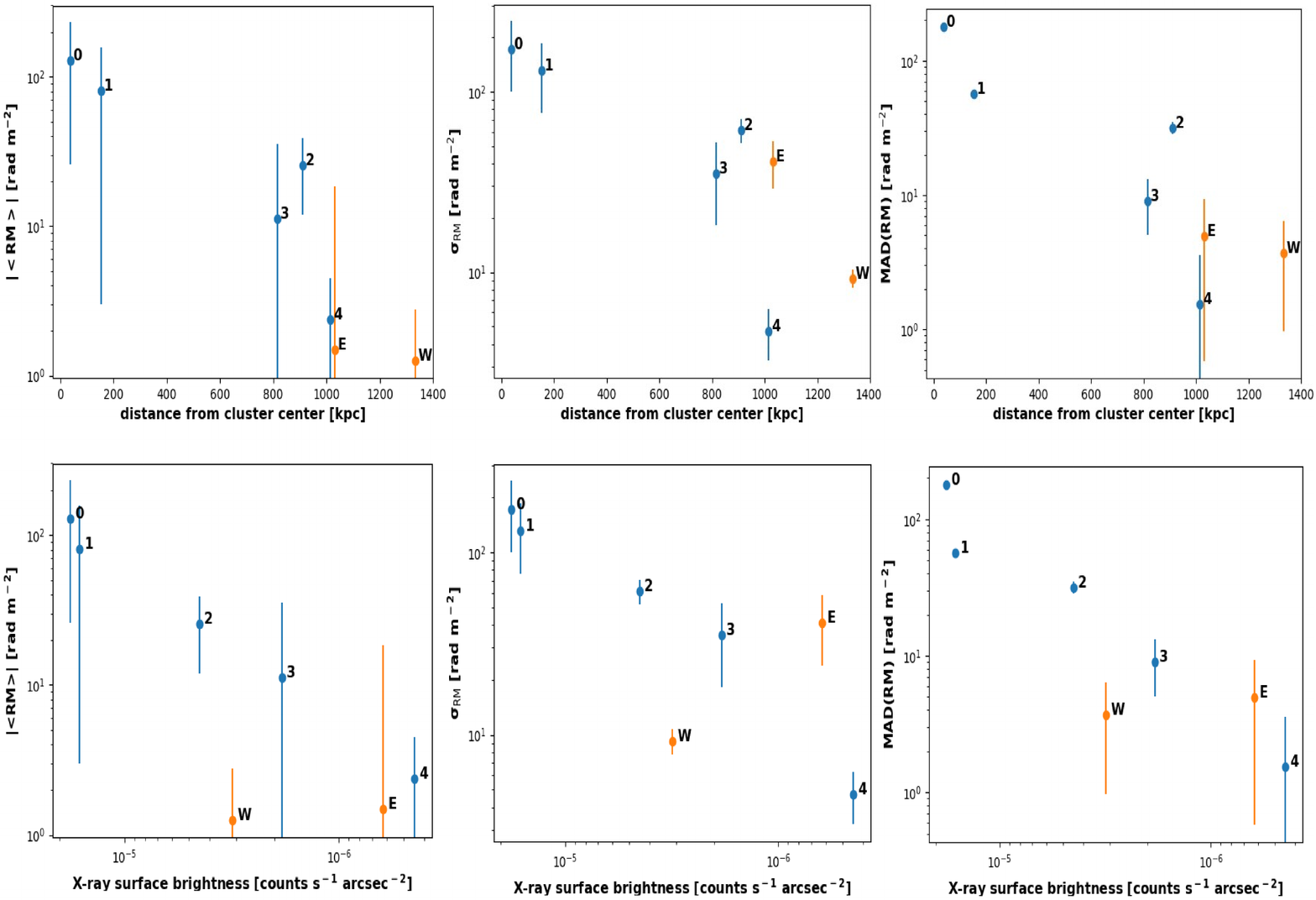}
    \caption{|<RM>|, $\sigma_\text{RM}$, and MAD(RM) of the sources in the cluster computed in the B-configuration plotted against the projected distance of each source from the X-ray peak (top panels) and against the X-ray surface brightness at the position of each source (bottom panels) . The uncertainties plotted for |<RM>| and $\sigma_\text{RM}$ are the $\pm1\sigma$ computed considering $n_\text{beam}$ independent samples for each source. The uncertainties plotted for MAD(RM) are derived from the median error on the single RM measurement, med($\sigma_{\phi}$). The five sources detected in polarization are numbered as in Fig.~\ref{fig:X} and the eastern and western relics are marked with the letters ‘‘E’’ and ‘‘W’’, respectively.}
    \label{fig:radial}
\end{figure*}

The radial profiles of the |<RM>|, $\sigma_\text{RM}$, and MAD(RM) values of the sources detected in polarization in the A2345 cluster are shown in the top panels of Fig.~\ref{fig:radial}. The radial distance of each source is computed as the projected distance between the X-ray peak and the brightest polarized pixel detected at the source position.

All the profiles clearly show a radial trend moving from the cluster center. This trend is expected if the Faraday rotation is mainly caused by the magneto-ionized medium of the cluster that produces a stronger effect on the sources seen in projection closer to the cluster center \citep[see, e.g.,][]{Bohringer16,Stasyszyn19}. As we noticed in Sec.~\ref{sec:Gal}, the Galactic contribution on the angular scales of the observed trend is expected to be negligible. The observed radial decrease of, both, RM and RM dispersion also disfavors the interpretation of the RM as due to the local environment of the radio sources. A layer of gas at the edge of the radio emitting plasma or in its close surroundings was proven to cause RM smaller than $\sim$20 rad m$^{-2}$ \citep[e.g.,][]{Guidetti12,Kaczmarek18}. Although a local contribution to the observed RM cannot be totally excluded, it is unlikely to be dominant over the ICM contribution \citep[see also,][]{Ensslin03b}. Furthermore, we observed Faraday-simple spectra which follows the expectations for an external Faraday screen. Following these considerations, we argue that the RM radial profile is likely to originate from the ICM, and thus that it can be used to infer the properties of the ICM magnetic field, as already done in previous work \citep[e.g.,][]{Murgia04,Bonafede10,Govoni17}.

Due to the complex X-ray morphology of A2345, the radial trend does not always follow a decrease in the X-ray surface brightness, and thus, of the thermal electron density integrated along the line-of-sight. This latter quantity is the real physical parameter we are interested in since it determines the amount of RM at the source position. Therefore, we also plotted the |<RM>|, $\sigma_\text{RM}$, and MAD(RM) values against the X-ray surface brightness measured at the position of each source (see bottom panels of Fig.~\ref{fig:radial}). The comparison between the radial and the surface brightness RM profiles is instructive because it shows that the spherical symmetry assumption does not hold for all regions of the cluster. While the E relic sector, with sources 0 and 4, is consistent with the spherical description, source 2 and the W relic have a local X-ray surface brightness which is not consistent with the radial dependence assuming spherical symmetry. We observe decreasing RM trends with decreasing X-ray surface brightness, as it is expected in the case the trend is caused by the decreasing column density of the ICM. Hence, these profiles can be used to constrain the properties of the ICM magnetic field using the RMs of all the detected polarized sources.

Among the shown profiles, we decided to focus on the median absolute deviation, MAD(RM). This choice is motivated by the fact that, in the simplest idealized model of ICM composed by cells of uniform size, equal thermal electron density, equal magnetic field strength and random orientation of the B vector, the RM dispersion is directly proportional to the cluster magnetic field \citep{Tribble91b}. Furthermore, we already noticed in Sec.~\ref{sec:Gal} that, while the mean RM observed in the external regions of the cluster can be attributed to the Galactic RM, the Milky way is not expected to contribute to the RM dispersion on the angular scales of the observed sources. The MAD is a good estimator for the RM dispersion and it is more resistant to outliers than $\sigma_\text{RM}$. Hence, this quantity will be compared with our simulated RM maps.

\section{Cluster magnetic field modeling}
\label{sec:Bmodel}

The determination of the cluster magnetic field properties from the RM measurements relies on the knowledge of, both, the thermal electron density and the magnetic field structure. In order to avoid simplistic assumptions, often used to solve the integral in Eq.~\ref{eq:RM}, we produced synthetic RM maps by taking into account realistic 3D models of the thermal electron density and of the magnetic field of a galaxy cluster. These RM maps can then directly be compared to observations, where the magnetic field model parameters can be constrained with a statistical approach.

This method has been proven to be successful for the study of the magnetic field in clusters \citep{Murgia04,Govoni06,Guidetti08,Bonafede10,Vacca12,Bonafede13,Govoni17}. However, to our knowledge, it has never been applied to the RM measurements of a radio relic. Only in \citet{Bonafede13} the RMs of seven sources seen in projection through the radio relic in the Coma cluster were used to probe the magnetic field properties in the relic and in the infall region. Using the RM of the relic itself can provide additional information on cluster magnetic fields.

Moreover, this is the first time in which this study is performed using the RM synthesis technique. The RM synthesis technique is in fact sensible to the internal Faraday rotation which is expected to be present in radio relics where layers of radio emitting plasma are mixed with the thermal gas \citep[see, e.g.,][]{Stuardi19}. The peak of the Faraday spectrum obtained at the relics is thus the sum of the polarized emission at each Faraday depth occupied by the emitting layers of the relic. In our case, we detected a Faraday-simple emission from the relics and this means that the single emitting layers are not resolved (see Sec.~\ref{sec:polrel} and Fig.~\ref{fig:pixels}). However, it is important to understand if the RM distribution observed at the relics can be used to probe the global magnetic field properties in the cluster. 

In this Section, we will apply this method to the RMs obtained from the central source 0, the more external source 4 and the E relic. Since the X-ray morphology of the cluster is strongly disturbed (see Sec.~\ref{sec:Xray}), a unique thermal electron density model, which would allow us to combine the RMs observed from all the sources and study their radial dependence, would be inaccurate. Hence, we decided to carry out the main analysis on the galaxy cluster region occupied by the E relic, using the thermal electron density profile obtained in this cluster sector. This is also the less disturbed region of the cluster, thus the assumption of the spherical symmetry is more suitable. We will reproduce the MAD(RM) profile and compare it to observations, in order to constrain the magnetic field profile from the center of the cluster up to the relic region. 

We will also study the MAD(RM) dependence from the X-ray surface brightness in order to be able to use the measurements obtained from all the sources together, with the expense of a larger uncertainty on the cluster geometry.

\subsection{Simulations of RM maps}
\label{sec:simul}

We used a modified version of the MiRo code described in \citet{Bonafede13}. We implemented important changes on the modeling of the magnetic field power spectrum, following recent results from cosmological MHD simulations \citep{Dominguez19}. The code firstly produces a mock 3D thermal electron density distribution based on X-ray observations. Then, it produces a 3D distribution of the  magnetic field, based on an analytical power spectrum within a fixed range of spatial scales. The magnetic field is scaled by the density profile and then normalized. Hence, the generated cluster magnetic field is tangled on both small and large scales, and it decreases radially. Finally, the code computes the cluster 2D RM map integrating the thermal electron density and magnetic field profile along one axis, solving Eq.~\ref{eq:RM}. We describe in more detail each of these steps.

\begin{enumerate}

\item \underline{The thermal electron density distribution} is built on the basis of the surface brightness profile derived in Sec.~\ref{sec:Xray}. For our statistical analysis we chose the sector of the E relic, i.e. between the position angles $160^{\circ}$ and $220^{\circ}$ (shown in Fig.~\ref{fig:X}), and we extracted the $n_{e}(r)$ profile following the double $\beta$-model (see Eq.~\ref{eq:2betadensity}). The six parameters of the double $\beta$-model were given as input in the simulation. 
As we noticed in Sec.~\ref{sec:Xray}, a single $\beta$-model is insufficient to describe the X-ray surface brightness in A2345. However, more complex models than a double $\beta$-model would not improve our results. We verified that the uncertainties on the fit parameters obtained from the double $\beta$-model result in smaller RM fluctuations with respect to fluctuations caused by the random nature of magnetic fields (see below).

Other input were the size of the simulated box and the pixel resolution (i.e., $\sim2^3 \ \text{Mpc$^3$}$ sampled with $512^3$ pixels of 4 kpc size). The center of the box was chosen to be the origin of the $n_{e}(r)$ profile as computed from the X-ray surface brightness peak.

\item \underline{The magnetic field power spectrum} is derived from the work of \citet{Dominguez19}. Using cosmological MHD simulations, the authors studied the evolution of the magnetic field in a set of highly resolved galaxy clusters. The authors found that the one-dimensional magnetic spectra of all the analyzed clusters can be well fitted to the same equation despite of the different cluster dynamical states:
\begin{equation}
    E_{B}(k) \propto k^{3/2} \bigg[ 1- \text{erf} \bigg(B\ln{\frac{k}{C}}  \bigg) \bigg] \ ,
\label{eq:Bpowerspec}
\end{equation}
where $k=\sqrt{ \sum_{i} k_i^2}$ (with $i$=1,2,3) is the wavenumber corresponding to the physical scale of the magnetic field fluctuations (i.e.,  $\Lambda\propto1/k$), $B$ is a parameter related to the width of the spectrum and $C$ is the wavenumber corresponding to the peak of the spectrum. Both $B$ and $C$ are found to depend on the dynamical state of the cluster while they only marginally depend on its mass \citep[see][for a discussion on those parameters]{Dominguez19}.

This parameterization allows us to use a more realistic power spectrum than those used in other work, where a Kolmogorov power-law spectrum is generally assumed \citep[e.g.,][]{Murgia04}. Indeed the turbulent dynamo, that is thought to be responsible for the amplification of the magnetic field in clusters, does not produce a power-law power spectrum for the magnetic field \citep[see, e.g.][for a recent review]{Schober15}. Instead, the slope of the power spectrum obtained from highly resolved MHD simulation is compatible with the Kazantsev model of dynamo for low wavenumbers, $E_{B}(k)\propto k^{3/2}$, and rapidly steepens from $\propto k^{-5/3}$ to $\propto k^{-2}$ or less after the peak of the spectrum \citep{Dominguez19}.

We used the $B$ and $C$ parameters of one of the merging clusters in the set at $z=0$ \footnote{See Tab.~1 in \citet{Dominguez19} and cluster with ID E5A.}. The fit is performed in the innermost $\sim2^3 \ \text{Mpc$^3$}$ region of the cluster using a $\sim512^3$ grid with a resolution of $\sim4 \ \text{kpc}$. This corresponds to a maximum fluctuation scale $\Lambda_\text{max}$ = 1 Mpc and a minimum scale $\Lambda_\text{min}$  $\sim$8 kpc. The parameters derived from the fit are $B$=1.054 and $C$ = 4.354 Mpc$^{-1}$ (corresponding to a power-spectrum peaking at $\sim$230 kpc). In our simulations we used the same box size, resolution and range of scales on which the fit was performed.

In order to obtain a divergence-free turbulent magnetic field, with the power spectrum described by Eq.~\ref{eq:Bpowerspec}, we first selected the corresponding power spectrum for the vector potential $\bf{\widetilde{A}(k)}$ in Fourier space $E_{A}(k) \propto k^{-2} E_{B}(k)$ \citep{Tribble91b,Murgia04}. For each pixel in Fourier space the amplitude, $A_{k,i}$, and the phase of each component of $\bf{\widetilde{A}(k)}$ are randomly drawn. $A=\sqrt{ \sum_{i} A_{k_i}^2 } $ is extracted from a Rayleigh distribution with scale parameter $E_{A}(k)$, while the phases are uniformly distributed between 0 and $2\pi$. The magnetic field vector in Fourier space is then ${\bf \widetilde{B}(k) = ik \times \widetilde{A}(k)}$ and has the desired power spectrum. $\bf{\widetilde{B}(k)}$ is transformed back into real space using an inverse Fast Fourier Transform algorithm. The resulting magnetic field, ${\bf B}$, has components $B_i$ following a Gaussian distribution, with $<B_i>=0$ and $\sigma_{B_i}^2=<B_i^2>$.

\item \underline{The radial profile of the magnitude of the magnetic field} is expected to scale with the thermal electron density \citep[e.g.,][]{Murgia04,Bonafede10}. A radial decrease of the magnetic field strength is also observed by MHD simulations \citep[e.g.,][]{Dolag99,Marinacci15,Vazza18,Dominguez19}. Therefore, we imposed that the cluster magnetic field scales with the thermal electron density following a power-law:
\begin{equation}
    |{\bf B}(r)| \propto n_e(r)^{\eta} \ ,
\end{equation}
where $\eta$ is a free parameter, as in \citet{Bonafede13}.

\item \underline{The normalization} of the magnetic field distribution is finally obtained imposing that the magnetic field averaged over the cluster volume (i.e., $\sim2^3 \ \text{Mpc$^3$}$) is $B_\text{mean}$. This is equivalent to fixing the value of $\sigma_{B_i}$. The value of $B_\text{mean}$ is the second parameter to be determined in the comparison with observations. This approach is slightly different from previous work where the normalization was performed fixing the average magnetic field value within the core radius or at the cluster center. This approach was preferred due to the greater complexity of the thermal electron density distribution found in A2345 with respect to other clusters. For comparison, we will also refer to the average magnetic field within the 200 kpc radius, $\langle B_0 \rangle$, computed over a set of ten random simulations having the same $B_\text{mean}$. Within ten simulations the value of $\langle B_0 \rangle$ has standard deviation below the $5\%$.

\end{enumerate}

Our magnetic field model considers a total of two free parameters, that can be finally determined comparing with our observations, namely $\eta$ and $B_\text{mean}$. The use of semi-analytical simulations including both the thermal electron density model obtained from the X-ray analysis and the power spectrum derived from MHD simulations give us the possibility to explore a wide range of magnetic field radial profiles.

Finally, we created a simulated RM map. The thermal electron density and the magnetic field along one axis (arbitrary chosen to be the $z$ axis of the cube) are numerically integrated according to Eq.~\ref{eq:RM}. The integration is performed from the center of the cluster, thus assuming that the sources and the relic lie on the plane parallel to the plane of the sky and crossing the cluster center. The resulting RM map has a size of $\sim2^2 \text{ Mpc$^2$}$ and a resolution of 4 kpc. The map is then convolved with a Gaussian kernel with FWHM equivalent to the restoring beam of the observed RM image (listed in Tab.~\ref{tab:pol}). 

\begin{figure}
    \includegraphics[width=1.\columnwidth]{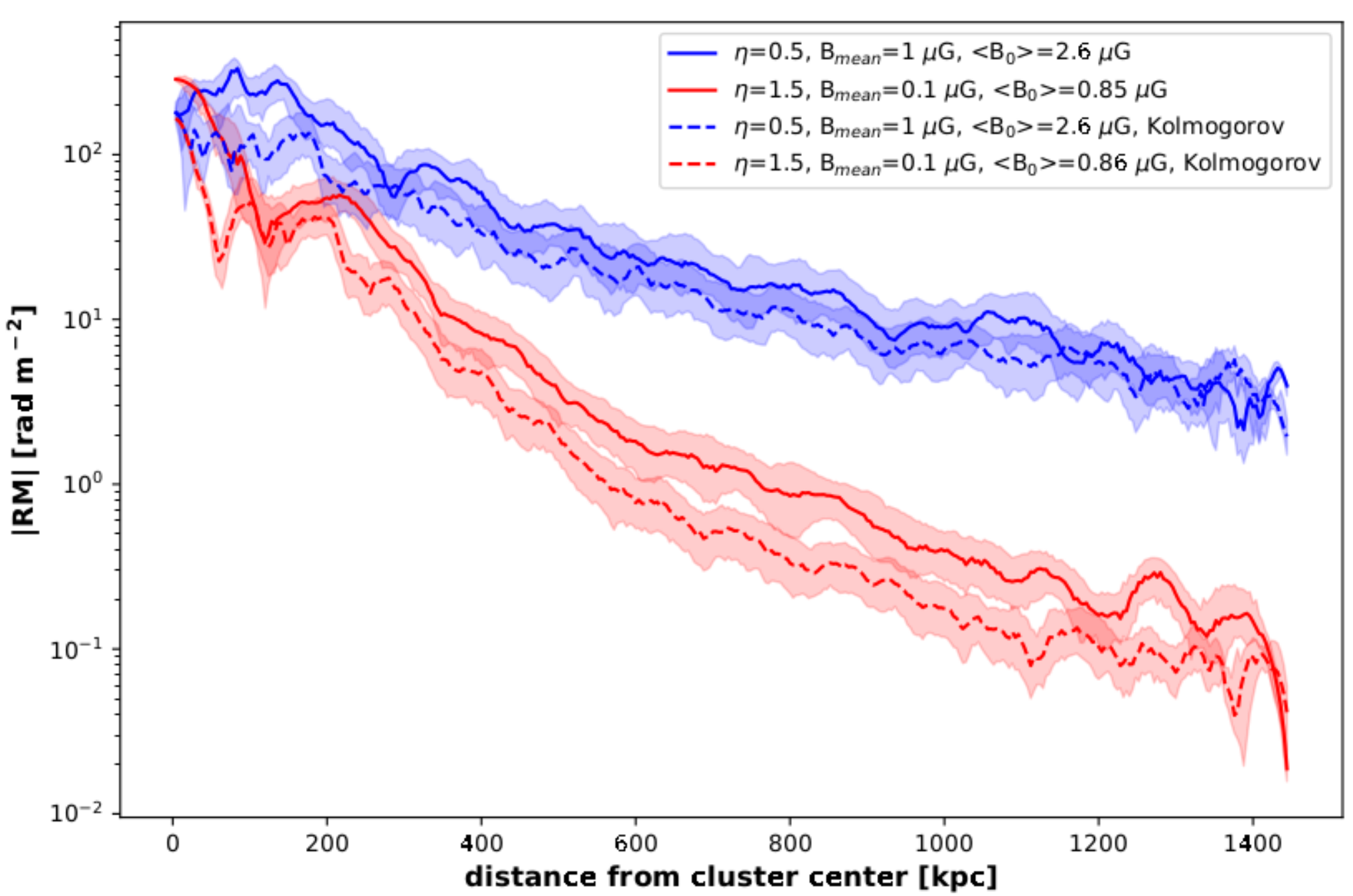}
    \caption{Comparison between RM profiles computed from different magnetic field models. The lines show the median |RM| profile with $35\%$ and $65\%$ boundaries computed from the simulated RM maps within annuli of increasing radius in a single random realization. The parameters of the models are listed in the label and the dashed lines refer to the same model computed with a magnetic Kolmogorov power spectrum.}
    \label{fig:RM_singlemodels}
\end{figure}

The RM profile can easily be computed from a single mock RM image considering annuli of increasing radius. As an example, we show the median RM profiles computed for two different combinations of $\eta$ and $B_\text{mean}$, namely [0.5,1] and [1.5,0.1], in Fig.~\ref{fig:RM_singlemodels}. The change of $\eta$ is responsible for a change in the slope of the RM profile while a change of $B_\text{mean}$ affects the overall normalization. We also compare these profiles with those obtained using a simple Kolmogorov power spectrum for the magnetic field. The Kolmogorov power spectrum is computed between $\Lambda_\text{min}$ = 8 kpc and $\Lambda_\text{max}$ = 230 kpc. In this way, the value of $\Lambda_\text{max}$ (where the Kolmogorov spectrum starts) coincides with the peak of the magnetic power spectrum derived from MHD simulations. We notice that, for the same magnetic field profile, the Kolmogorov power spectrum produces, on average, lower values of median RM by a factor $\sim$2. This means that the same observational RM radial trend would be fitted with a higher magnetic field for a power-law Kolmogorov spectrum, with respect to the one that would be fitted by our model. Considering the model with $B_\text{mean}$ = 0.1 $\mu$G, the RM profile obtained with the Kolmogorov power spectrum appears steeper: in this case, a fit performed with this model would underestimate the $\eta$ parameter. This comparison confirms that, in order to derive detailed cluster magnetic fields properties, it is essential to use a more realistic magnetic power spectra (see Sec.~\ref{sec:kolmog}).

Given the random nature of the magnetic field distribution, the RM and RM dispersion in a certain position of the cluster vary depending on the initial random seed of the simulation for different realizations of the same model. To better compare the observed and simulated quantities, the RM image can be also clipped at the distance of a source from the cluster X-ray peak and blanked following the shape of the given source. Hence, the same observational sampling bias is introduced in the simulated quantities.

\subsection{Constraining magnetic field properties}
\label{sec:comparison}

In order to asses the best match between observation and simulations, we build a set of simulations varying $B_\text{mean} = 0.05,0.1,0.5,1 \ \mu\text{G}$ and $\eta = 0.5,1,1.5,2$. For each combination of the two parameters we build ten realization starting from different random seeds. The RM maps were convolved with a FWHM of 24 kpc corresponding to the 8$\arcsec$ resolution of the B-configuration observation. From each simulation we extracted the mock RM image at the distance and with the shape of each source, as described in Sec.~\ref{sec:simul}.

\begin{figure}
    \includegraphics[width=1.\columnwidth]{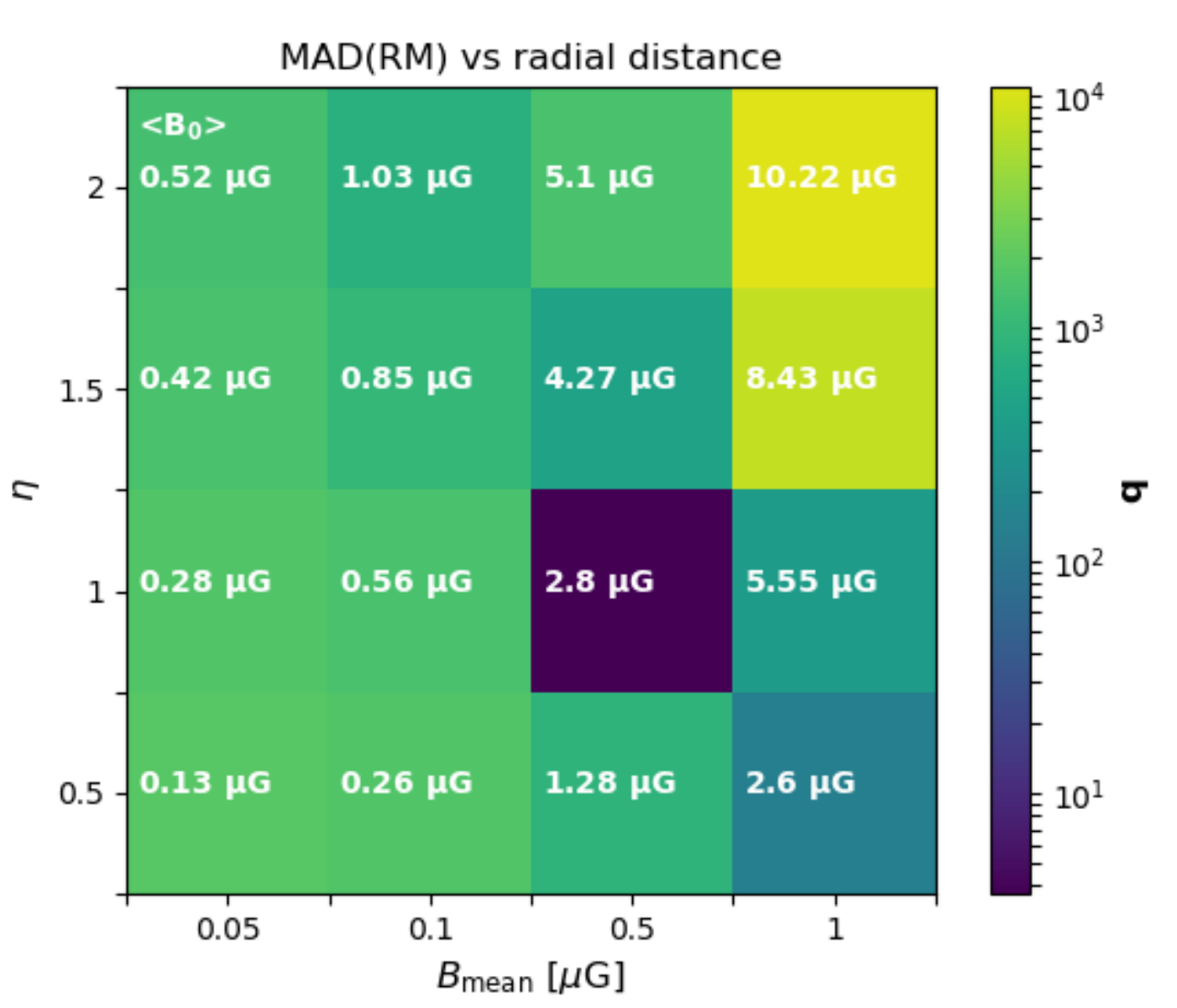}
    \includegraphics[width=1.\columnwidth]{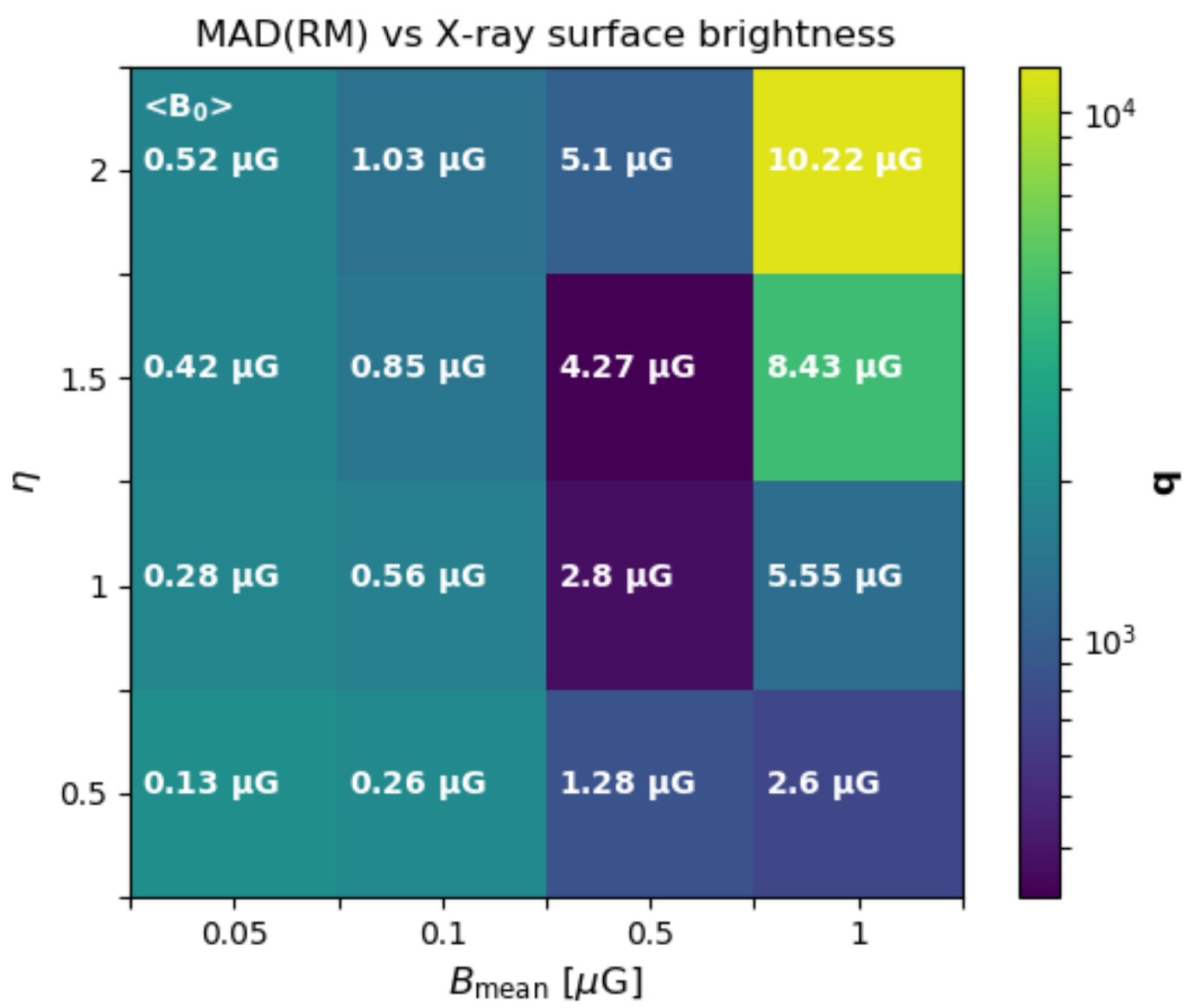}
    \caption{Plots of the $q$ statistics derived for the radial MAD(RM) profile (upper panel) and for the profile against X-ray surface brightness (bottom panel) for several combinations of the model parameters $\eta$ and $B_\text{mean}$. For each model we show the value of the average magnetic field computed for ten different realizations in the central volume of the cluster within 200 kpc radius.}
    \label{fig:q_models}
\end{figure}

As explained in Sec.~\ref{sec:RMprof}, we decided to carry out the comparison between observation and simulation using the values of the median absolute deviation, MAD(RM). In this case the best match with observation is obtained for the minimum of the quantity:
\begin{equation}
    q=\sum_{i=0,4,\text{E}} \Big( {\frac{\rm MAD(RM)_{i,obs} - \langle MAD(RM)_{i,sim} \rangle}{\rm errMAD(RM)_{i,obs}}} \Big)^2 \ ,
    \label{eq:q_parameter}
\end{equation}
where $i$ = 0, 4, E refers to the three sources and the average is computed over the ten different realizations of the same magnetic field model. The error on the observed MAD is computed as errMAD(RM)$_\text{i,obs}$ = $\sqrt{2}\text{med}(\sigma_{\phi})$ (the MAD is the difference between two single RM estimates which are affected by the same median error). The resulting $q$ parameters for the explored combinations of $\eta$ and $B_\text{mean}$ are shown in the top panel of Fig.~\ref{fig:q_models}. 

The minimum is reached for $\eta$=1 and $B_\text{mean}$ = 0.5 $\mu$G. This magnetic field model has an average central magnetic field $\langle B_0 \rangle$ = 2.8$\pm$0.1 $\mu$G (where the average is computed over the ten random realizations and the uncertainty is the standard deviation). The average magnetic field at the relic (i.e., computed in a spherical shell of 200 kpc radius at a distance of 1 Mpc from the center) is $\sim$0.3 $\mu$G.

\begin{figure}
    \includegraphics[width=1\columnwidth]{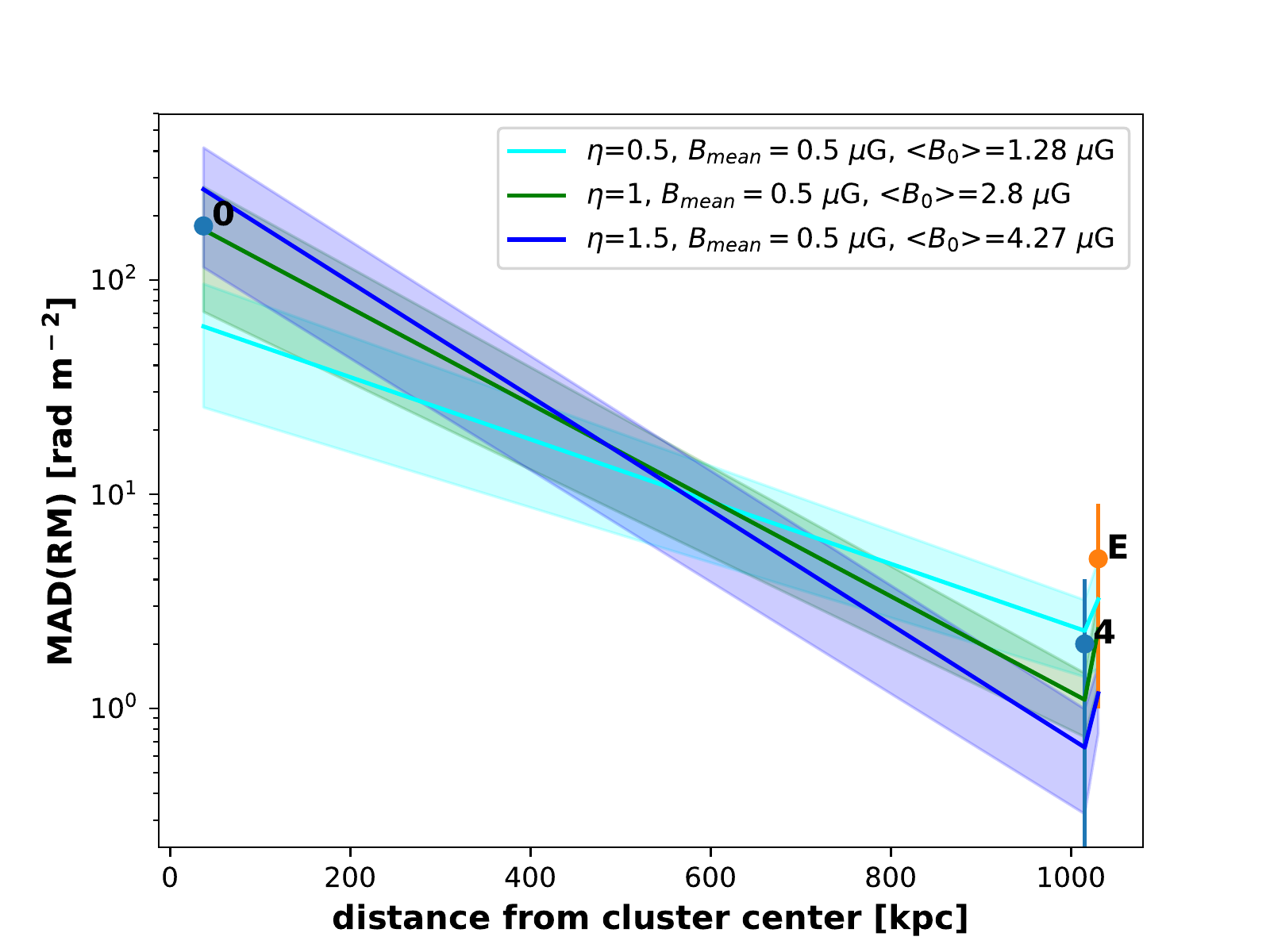}
    \includegraphics[width=1\columnwidth]{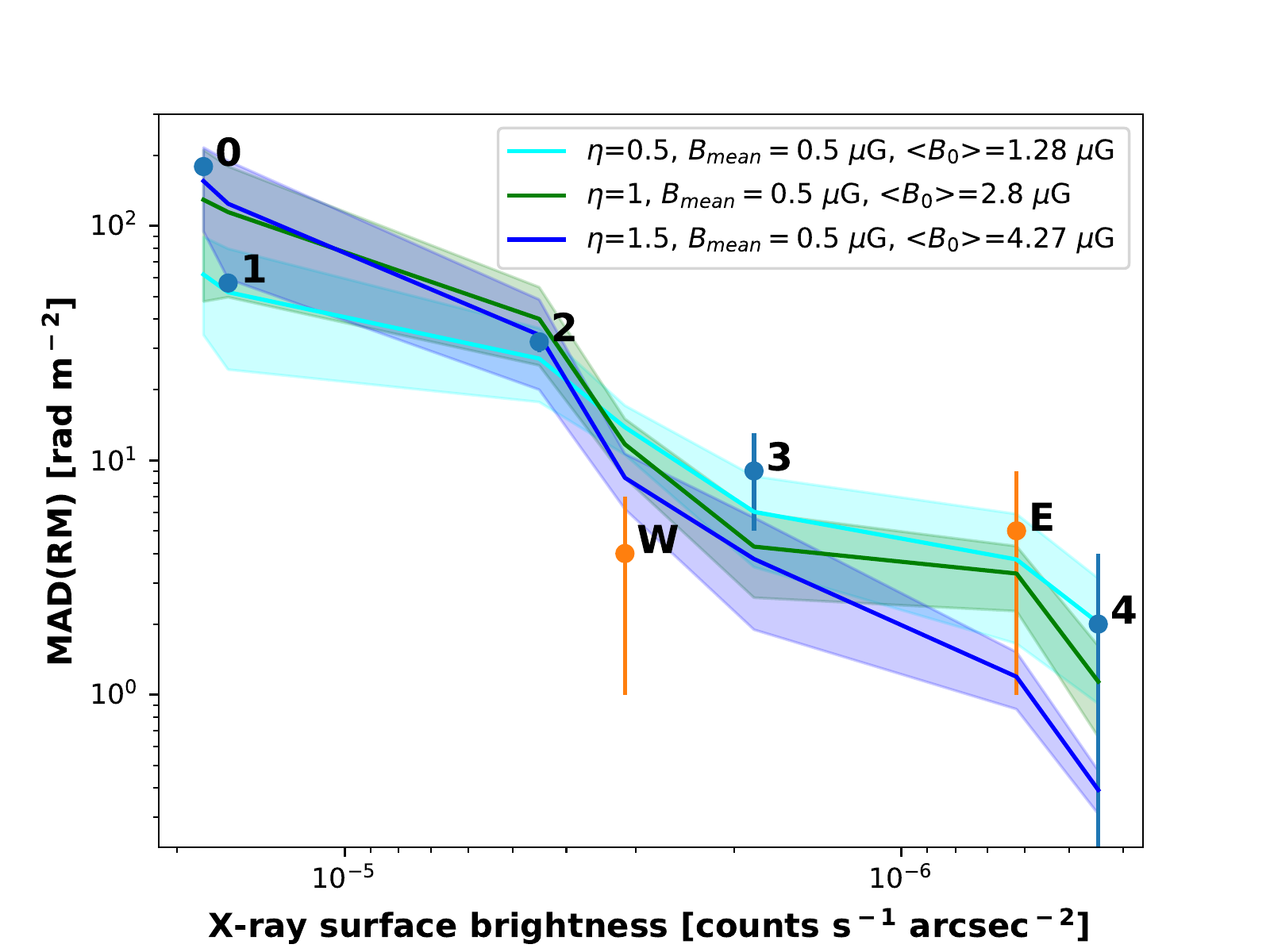}
    \caption{Simulated MAD(RM) profiles compared with observations. Lines show the average obtained from ten different realizations of the same model with shadowed areas showing the standard deviation. In the sector containing the E relic the profiles are plotted against the radial distance from the X-ray peak (top panel). The profiles computed using all the detected sources are plotted against the X-ray surface brightness (bottom panel), as explained in Sec.~\ref{sec:comparison}. The best model is the green one.}
    \label{fig:RM_meanmodels}
\end{figure}

The best MAD(RM) profile derived from simulations is compared with observed values in the top panel of Fig.~\ref{fig:RM_meanmodels}. In the same plot, we also show two simulated MAD(RM) profiles obtained with the same $B_\text{mean}$ (i.e., 0.5 $\mu$G) but with different $\eta$.

We repeated the same test on $q$ including in the profile the E relic observation performed with the C array. In this case the maps were convolved with a FWHM of 96 kpc, corresponding to the 30.5$\arcsec$ resolution beam. The results (not shown here) are very similar to the results obtained without including this observation, and they constrain the same magnetic field model.

As we noticed in Sec.~\ref{sec:RMprof}, each sector of A2345 shows a different X-ray surface brightness profile and thus a different underlying density distribution. Therefore, it is not possible to fit the same radial profile including all the sources. Instead, it is possible to exploit the dependence of MAD(RM) from the observed X-ray surface brightness (see Fig.~\ref{fig:radial}). Independently of the underlying thermal gas density distribution, the X-ray surface brightness observed at the position of each source is a good proxy for the thermal electron column density at that position. A large uncertainty is represented by the unknown position of each source along the line-of-sight within the X-ray emitting volume.

In order to test our model with a larger number of observational points, we extracted the simulated RM images of each source at the radial distance derived from the observed X-ray surface brightness, i.e. inverting Eq.~\ref{eq:2betasb}. This method allows us to enlarge the statistics, at the expense of a larger uncertainty in the adopted density model. We computed the new values of MAD(RM) for all the combination of $B_\text{mean} = 0.05,0.1,0.5,1 \ \mu\text{G}$ and $\eta = 0.5,1,1.5,2$ and we computed the $q$ parameter (see Eq.~\ref{eq:q_parameter}), with $i$ = 0, 1, 2, 3, 4, E, W. In this case, the best-fitting model is the one with $\eta=1.5$ and $B_\text{mean}$ = 0.5 $\mu$G but a very similar $q$ value is obtained for $\eta=1$ (see bottom panel of Fig.~\ref{fig:q_models}). 

The three models with $B_\text{mean}$ = 0.5 $\mu$G and $\eta$ = 0.5, 1, 1.5 are compared with the observed MAD(RM) profile plotted against the X-ray surface brightness in the bottom panel of Fig.~\ref{fig:RM_meanmodels}. Although the minimum $q$ is obtained for the model with $\eta$ = 1.5, this seems to be mainly due to the MAD(RM) value of the central sources while peripheral sources are better described by the model with $\eta$ = 1. This confirms that this latter model better describes the magnetic field profile in the radio relic sector, and that the same magnetic field profile is able to reasonably reproduce the RMs observed in the entire cluster.

To summarize, we found that a magnetic field tangled on scales between 8 and 1000 kpc, following a power spectrum defined by Eq.~\ref{eq:Bpowerspec} with a peak at $\sim$230 kpc, best describes our data with a central magnetic field $\langle B_0 \rangle$ = 2.8$\pm$0.1 $\mu$G and $\eta\sim1$. The average magnetic field at the position of the E relic is thus constrained to be $\sim$ 0.3 $\mu$G.

It is necessary to notice that our simulations assume spherical symmetry and that the RM computation further assumes that all the observed sources and relic are aligned on the same plane. A recent work by \citet{Johnson20}, identified these assumptions as one of the principal uncertainties on the determination of cluster magnetic fields from Faraday rotation measurements. The authors stated that RM-estimated central magnetic field strengths suffer for an uncertainty of a factor $\sim$3 due to the, still, unknown parameters of the the model used to interpret RM measurements. In our case, the assumption on the position of sources 0, 1, 2, and of the E relic is well motivated by the work by \citet{Boschin10} which obtained the redshifts of the sources and found that the merger that originated the E relic has its main component on the plane of the sky. This is not valid for the other two sources and for the W relic. However, the analytical expression often used to derive the magnetic field strength from RM dispersion in galaxy clusters \citep[][Eq. 3]{Felten96} allows us to state that the uncertainty on the position of the sources in the cluster can account for a factor of $\sqrt{2}$ uncertainty on our magnetic field estimates. This uncertainty cannot be avoided even with the use of numerical simulations. We note that this is not the dominant source of errors given the assumptions we have to make to derive the magnetic field estimate.

Another source of uncertainty in our modeling is introduced by the assumption that the ICM magnetic field strength follows a Maxwellian distribution. In fact, cosmological MHD simulations demonstrated that the 3D magnetic field distribution shows strong departures from a simple Maxwellian distribution and that this may have a strong impact on the RM-based estimate of the central magnetic field strength \citep{Vazza18}. In order to verify this hypothesis, we would need RM information from a larger fraction of the sky area cover by the cluster. With the lack of the necessary statistic, the assumption of a magnetic field distribution other than the Gaussian would only add more free parameters to our model. Taking note of these considerations, it is clear that the uncertainty on the value of $\langle B_0 \rangle$ = 2.8 $\mu$G is larger than the one derived from the standard deviation between the ten realizations of the same model. 

\subsection{Comparison with a Kolmogorov power spectrum}
\label{sec:kolmog}

\begin{figure}
    \includegraphics[width=1.\columnwidth]{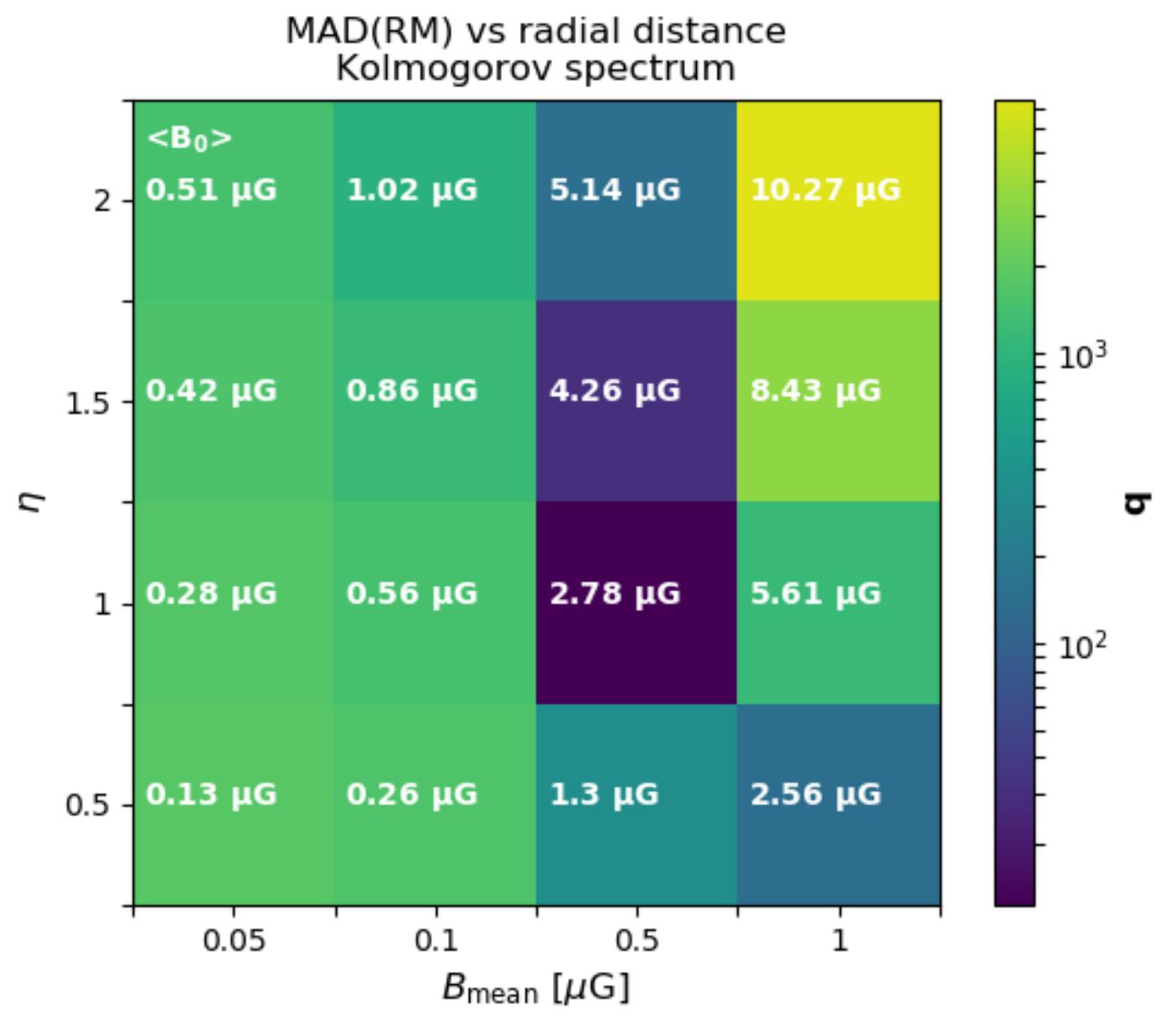}
    \includegraphics[width=1.\columnwidth]{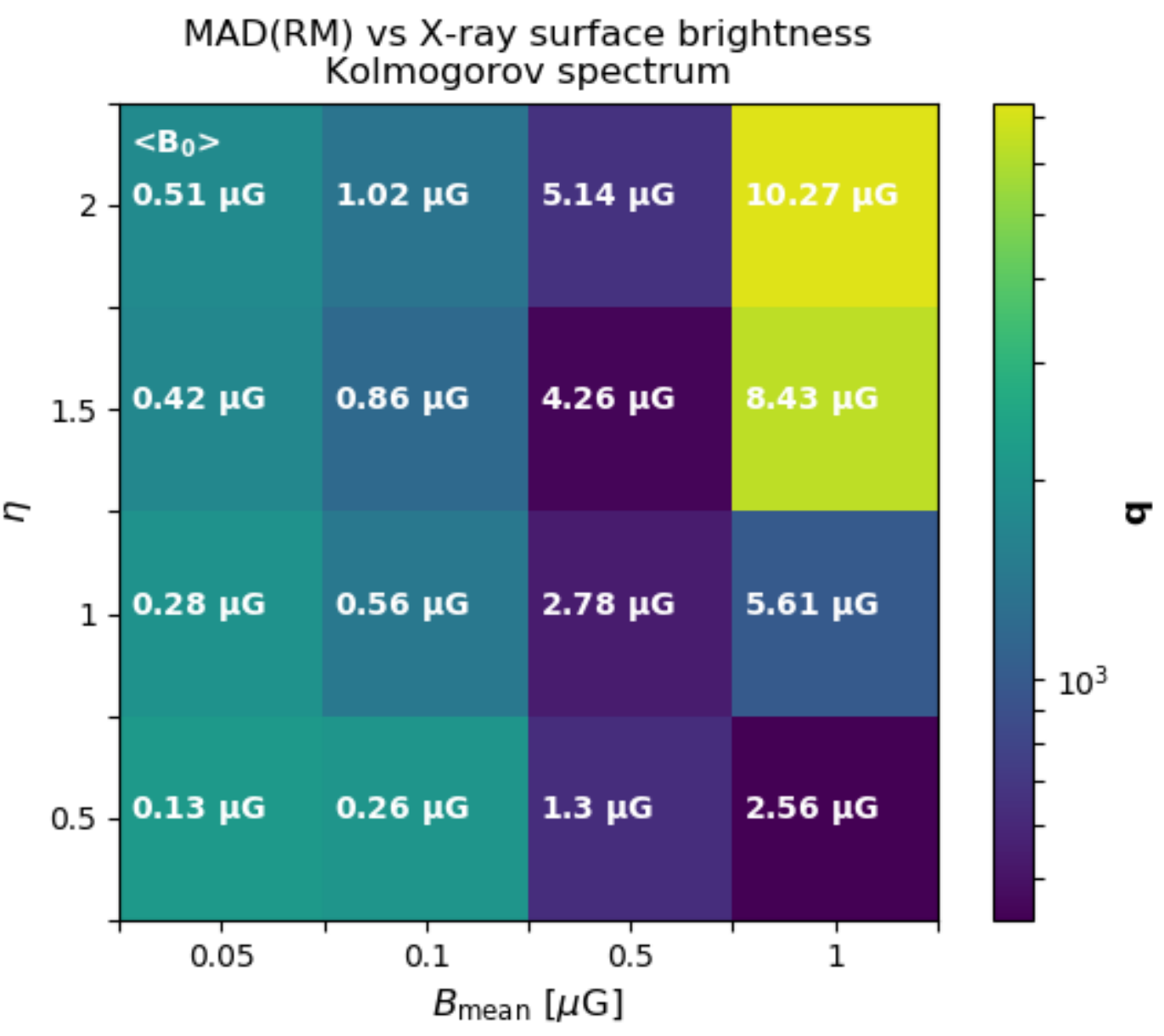}
    \caption{Plots of the $q$ statistics derived for the radial MAD(RM) profile (upper panel) and for the profile against X-ray surface brightness (bottom panel) for several combinations of the model parameters $\eta$ and $B_\text{mean}$. For each model we show the value of the average magnetic field computed for ten different realizations in the central volume of the cluster within 200 kpc radius.}
    \label{fig:q_models_kolmog}
\end{figure}

In this Section, we repeated the tests performed in Sec.~\ref{sec:comparison} using a magnetic field model with a Kolmogorov power spectrum. This was done to further investigate the differences caused by the use of a realistic power spectrum derived from MHD simulations, with respect to previous work which adopted a power law spectrum to describe magnetic field fluctuations. To build mock RM maps we repeated the four steps described in Sec.~\ref{sec:simul} but this time we imposed a Kolmogorov power spectrum having components $|B_{k}|^2 \propto k^{-11/3}$ in the Fourier space. As done for the RM profiles shown in Fig.~\ref{fig:RM_singlemodels}, the Kolmogorov power spectrum is computed between $\Lambda_\text{min}$ = 8 kpc and $\Lambda_\text{max}$ = 230 kpc. With this arbitrary choice we decided to use the Kolmogorov spectrum which is more similar to the one obtained from MHD simulations since $\Lambda_\text{max}$ coincides with the peak of the magnetic power spectrum described by Eq.~\ref{eq:Bpowerspec}. With a different choice of scales (based for example on the scale of the RM fluctuations observed from single sources) we could of course obtain different results. We tested the same sets of $B_\text{mean}$ and $\eta$ parameters and we computed the $q$ statistics for the radial and X-ray surface brightness RM profiles as done in Sec.~\ref{sec:comparison}. 

The results are shown in Fig.~\ref{fig:q_models_kolmog}. When comparing the MAD(RM) radial profile computed for sources 0, 4, and the E relic with the simulated one (top panel), the preferred model is the same obtained in Sec.~\ref{sec:comparison} (i.e., the one with $B_\text{mean}$ = 0.5 $\mu$G and $\eta\sim1$). However, we notice that a similar value of $q$ is obtained also for $\eta\sim1.5$ which implies a higher central magnetic field, $\langle B_0 \rangle$ = 4.26$\pm$0.1 $\mu$G. A finer sampling of the $B_\text{mean}$ and $\eta$ parameter space would lead to different results for the two models having different magnetic field power spectra. Considering the MAD(RM) profile versus the X-ray surface brightness for all the polarized sources detected in A2345 (bottom panel of Fig.~\ref{fig:q_models_kolmog}), the minimum of the $q$ parameter is instead reached for the model with $B_\text{mean}$ = 1 $\mu$G and $\eta\sim0.5$. The central magnetic field of this model, $\langle B_0 \rangle$ = 2.78$\pm$0.1 $\mu$G, is very similar to the one obtained in Sec.~\ref{sec:comparison} (i.e., 2.8$\pm$0.1 $\mu$G) but the magnetic field radial profile has a flatter power-law index. Also in this case, a low values of $q$ is also obtained for $\eta\sim1.5$ and $\langle B_0 \rangle$ = 4.26$\pm$0.1 $\mu$G. 

Overall these results are in agreement with the differences observed between the RM profiles shown in Fig.~\ref{fig:RM_singlemodels}. The models using a Kolmogorov spectrum for magnetic field fluctuations tend to lead to magnetic fields with higher strength, or shallower radial dependence, than models with a power spectrum derived from recent MHD simulations. We note that the maximum scale of magnetic field fluctuations used in the Kolmogorov spectrum was decided on the basis of MHD simulations because, due to the small extent of detected polarized sources, this information could not be derived directly from RM data. Therefore, these simulations are as similar as possible to the simulations obtained with the spectrum derived from Eq.~\ref{eq:Bpowerspec}. The detailed investigation of the impact of a different power spectrum on magnetic field estimates may be the subject for further studies.

\section{Discussion}
\label{sec:discus}

Under the assumption that the RM and $\sigma_\text{RM}$ radial profiles observed in the A2345 galaxy cluster are dominated by Faraday rotation in the ICM, we constrained the magnetic field profiles that, within the framework of our model, may better reproduce the observations.

Several statistical studies demonstrated that the Faraday rotation of sources seen in projection within clusters decreased with the radial distance from the cluster center \citep[e.g.,][]{Clarke01,Johnston-Hollitt04,Bohringer16,Stasyszyn19}. Fewer of these kinds of analyses were performed on single clusters, since current facilities allow the detection of few polarized sources per square degree \citep[see, e.g.,][]{Rudnick14}. The RM radial trend we observed in the A2345 galaxy cluster is a single-cluster confirmation of previous statistical studies, as was also found in Abell 514 \citep{Govoni01b}

One of the first attempts to unveil the magnetic field profile and power spectrum of a single cluster was performed by \citet{Murgia04}, who used RMs from three galaxies observed within Abell 119. Similar work was performed on Abell 2255, Abell 2382 and Abell 194 \citep{Govoni06,Guidetti08,Govoni17}. Other studies were performed exploiting the presence of a central radio halo or a single extended polarized radio source observed at high angular resolution \citep{Vacca10,Vacca12}. Another notable exception is the Coma galaxy cluster that, thanks to its proximity, spans more than one degree in projected size. Its intra-cluster magnetic field was studied with great detail using the RMs of 14 radio galaxies and with a method similar to the one we adopted in this paper \citep{Bonafede10,Bonafede13}. In this work, the intra-cluster magnetic field was described with a Kolmogorov power spectrum on scales between 2 and 34 kpc. The best-fit parameters were found to be $\langle B_0 \rangle$ = 4.7 $\mu$G and $\eta$ = 0.5. The authors also inferred that the magnetic field should be amplified by a factor of $\sim3$ throughout the entire merging region where the Coma radio relic is observed. 

\begin{table*}
    \centering
\caption{Comparison of results in literature. Column 1: galaxy cluster name; Column 2: mass estimate within $r_{500}$. All the estimates refer to the hydrostatic mass from Sunyaev-Zeldovich effect observations \citep{Planck16b}, apart from the poor galaxy cluster Abell 194 for which we used the X-ray mass estimate from \citet{Lovisari15}; Column 3: redshift; Column 4: dynamical state of the cluster based on literature search. When the classification is uncertain a ‘‘(?)’’ symbol is used; Column 5: magnetic field power spectrum index in the expression $|B_{k}|^2 \propto k^{-n}$. In this work the magnetic field power spectrum is assumed as explained in Sec~\ref{sec:simul}; Column 6 and 7: minimum and maximum scale of the magnetic field power spectrum fluctuations; Column 8: average magnetic field at the cluster center; Column 9: average magnetic field in a $\sim$1 Mpc$^3$ volume; Column 10: radial slope of the magnetic field profile; Column 11: Reference paper. A value is marked with an asterisk when it is assumed and fixed in the model rather than derived from observed parameters.}
\begin{tabular}{|c|c|c|c|c|c|c|c|c|c|c|}
\hline
  \multicolumn{1}{|c|}{Galaxy cluster} &
  \multicolumn{1}{|c|}{$M_{500}$} &
  \multicolumn{1}{c|}{$z$} &
  \multicolumn{1}{c|}{Dynamical state} &
  \multicolumn{1}{c|}{n} &
  \multicolumn{1}{c|}{$\Lambda_\text{min}$} &
  \multicolumn{1}{c|}{$\Lambda_\text{max}$} &
  \multicolumn{1}{|c|}{$\langle B_0 \rangle$} &
  \multicolumn{1}{|c|}{$\langle B_\text{1Mpc$^3$} \rangle$} &
  \multicolumn{1}{c|}{$\eta$} &
  \multicolumn{1}{c|}{Ref.} \\
  \multicolumn{1}{|c|}{} &
  \multicolumn{1}{|c|}{($10^{14} M_{\odot}$)} &
  \multicolumn{1}{c|}{} &
  \multicolumn{1}{c|}{} &
  \multicolumn{1}{c|}{} &
  \multicolumn{1}{c|}{(kpc)} &
  \multicolumn{1}{c|}{(kpc)} &
  \multicolumn{1}{c|}{($\mu$G)} &
  \multicolumn{1}{c|}{($\mu$G)} &
  \multicolumn{1}{c|}{} &
  \multicolumn{1}{c|}{} \\
\hline
  Abell 119 & 3.4 & 0.04 & merging & 2 & 6$^{*}$ & 770$^{*}$ & 5 & 1.5 & 0.9 & \citet{Murgia04} \\
  Abell 2255 & 5.4 & 0.08 & merging & 2-4 & 4$^{*}$ & 512$^{*}$ & 2.5 & 1.2 & 0.5$^{*}$  & \citet{Govoni06} \\
  Abell 2382 & 2.0 & 0.06 & relaxed (?) & 11/3 & 6$^{*}$ & 35 & 3.3 & 1 & 0.5$^{*}$ & \citet{Guidetti08} \\    
  Coma & 7.2 & 0.02 & merging & 11/3 & 2 & 34 & 4.7 & 2 & 0.5 & \citet{Bonafede10} \\ 
  Abell 665 & 8.9 & 0.18 & merging & 11/3$^{*}$ & 2$^{*}$ & 34 & 1.3 & 0.75 & 0.5$^*$ & \citet{Vacca10} \\
  Abell 2199 & 2.9 & 0.03 & relaxed & 2.8 & 0.7 & 35 & 11.7 & 0.2 & 0.9 & \citet{Vacca12} \\
  Abell 194 & 0.3 & 0.02 & relaxed & 11/3$^{*}$ & 1$^{*}$ & 64 & 1.5 & 0.3 & 1.1 & \citet{Govoni17} \\
  Abell 2345 & 5.9 & 0.18 & merging & Eq.~\ref{eq:Bpowerspec} & 8$^{*}$ & 1000$^{*}$ & 2.8 & 1.2 & 1.0 & This work \\

\hline\end{tabular}
\label{tab:literature}
\end{table*}

\begin{figure}
    \includegraphics[width=1.\columnwidth]{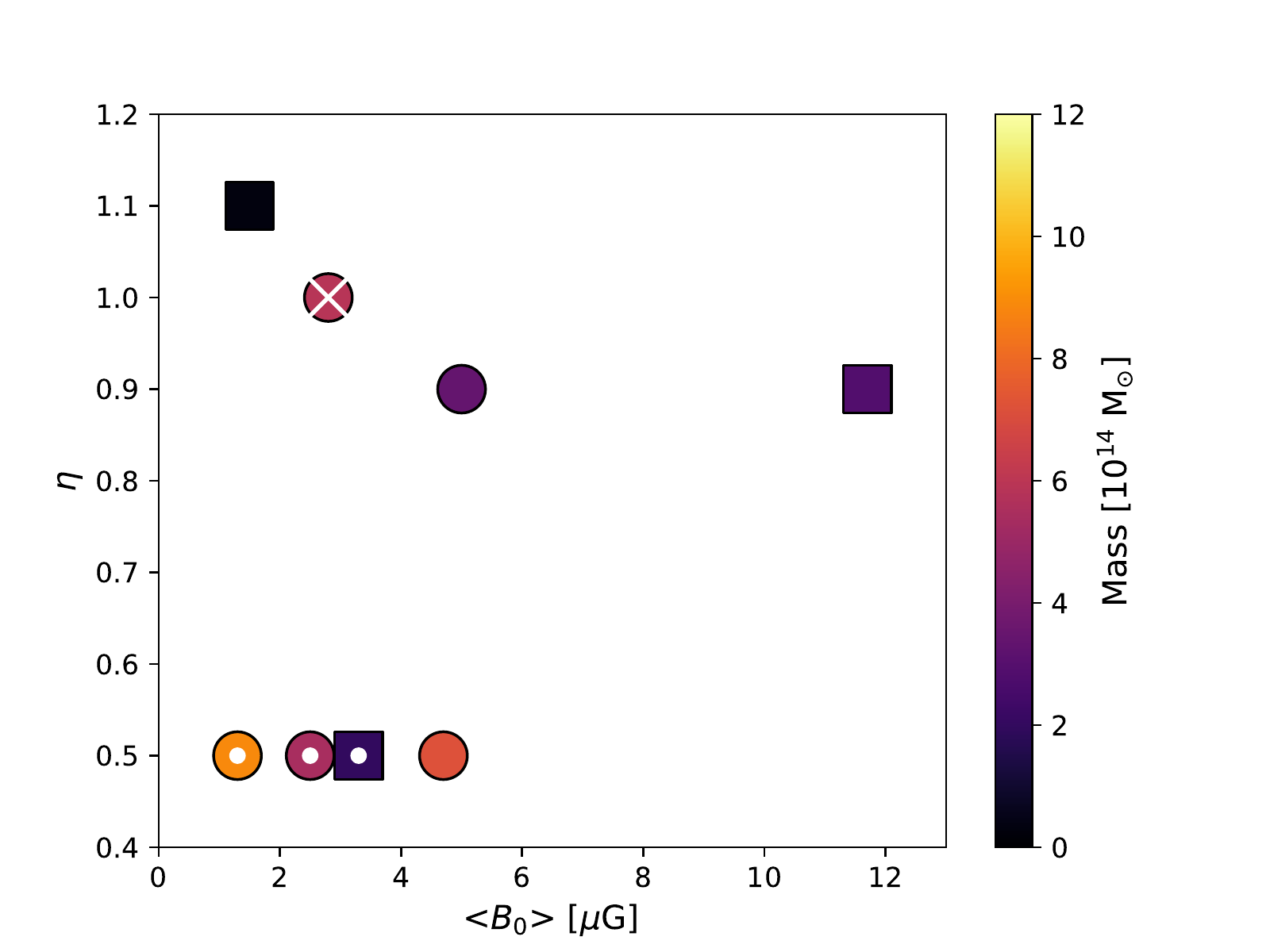}
    \caption{Comparison with $\eta$ and B$_0$ values found in the literature. Square markers show relaxed clusters while circles show merging clusters. The value obtained in this work is marked with an ‘‘X’’ while clusters for which the value of $\eta$ = 0.5 was assumed  a priori are marked with a filled white circle. The values are listed in Tab.~\ref{tab:literature}.}
    \label{fig:literature}
\end{figure}

We list the main results of the aforementioned work in Tab.~\ref{tab:literature} and we include the results obtained in Sec.~\ref{sec:comparison}. We computed the average magnetic field strength in the central $\sim$1 Mpc$^3$ of the simulated cube. This value is only computed for reference, since in our case we only modeled a sector of the cluster. The parameters of the magnetic field profile obtained by each work are also plotted in Fig.~\ref{fig:literature}.

The values obtained for $\langle B_0 \rangle$ range between 1.3 and 11.7 $\mu$G and do not correlate with the mass of the galaxy cluster. The value of $\eta$ we obtained in this work agrees with the literature. In most of the previous work, the value of $\eta$ = 0.5 was assumed, and only in the case of Coma it was derived from a comparison with observations. In fact, if the magnetic field strength decreases as the square root of the thermal electron density, the gas is at the equilibrium since the magnetic energy density decreases as the gas energy density. Higher values of $\eta$ lead to a higher central magnetic field and to a stronger radial decrease of the magnetic energy density. However, it shall be stressed that the $\langle B_0 \rangle$ and $\eta$ parameters reported from the literature were derived with rather varied approaches to the 3D modeling of magnetic fields, and, in particular, our work here is the first to assume a power spectrum that departs from a simple power-law, in agreement with small-scale dynamo simulations. As shown in Sec.~\ref{sec:kolmog}, this can have an impact on the magnetic field parameters derived by comparing to simulations. Furthermore, the largest scale of our spectrum, derived from MHD simulations, is 1 Mpc, which is $\sim$30 times larger than the largest scale obtained for the power spectrum in the Coma cluster, and has a peak at $\sim$230 kpc. Also the physical condition of the galaxy cluster can play a role since it is still not clear if the more massive and relaxed clusters have a larger central magnetic field with respect to merging systems \citep{vanWeeren19,Stasyszyn19}. 

In our magnetic field model we assumed a unique magnetic field power spectrum to describe the entire volume of the cluster. In particular, the power spectrum was retrieved from the cosmological MHD simulation of a merging galaxy cluster \citep{Dominguez19}. However, it is possible that existing shocks change the magnetic power spectrum. This would be a possible scenario for the observed relic. Recently, \citet{Dominguez20} studied the impact of shocks on the magnetic power spectrum. This study used MHD simulations of Mach number 2-3 shocks propagating through a turbulent ICM 200$^3$ kpc$^3$ box. In this work, the authors concluded that the turbulence created after the shock passage may have an impact of the local magnetic field power spectrum. In particular, after the shock passage, the power spectrum shifts the power spectrum on physical scales $\gtrsim$ 50 kpc to larger scales (i.e, lower wave-numbers) while leaving scales below 10 kpc largely unaffected. In this case, the intra-cluster magnetic field profile would be best represented by the RM dispersion profile since this is determined by magnetic field fluctuation on scales smaller than the sources size. Furthermore, a global power spectrum model may not be sufficient to describe the magnetic field profile in the entire cluster when a merger is occurring, as was also pointed out by \citet{Govoni06}.

We also obtained an estimate of the magnetic field strength at the E relic. Assuming equipartition \citet{Bonafede09a} obtained an estimate of 0.8 $\mu$G for the E relic. This value is 2.7 times larger than the one that we obtained for the model with $B_\text{mean}$=0.5 $\mu$G (i.e., 0.3 $\mu$G). This discrepancy can be motivated by the large number of assumptions that should be taken into account in the equipartition estimate and that could have lead to an overestimation of the magnetic field. In any case, no physical reason for relics to be at the equipartition exists. On the other hand, it is also possible that projection effects play a role and that the RMs we obtained from the relic only sample the intra-cluster medium outside a thin shell in front of the relic. In this case, our magnetic field would be underestimated. Another important source of uncertainty is the assumption of spherical symmetry in the determination of the electron density profile. A discrepancy between the magnetic field values obtained from the equipartition estimate and with the RM analysis in radio relics was already observed \citep[e.g.,][]{Ozawa15}. 

No evident RM jump was found at the position of the E relic, as for the Coma radio relic. In any case, with the current modeling we cannot investigate if magnetic field amplification occurs in the relic region, as found for the Coma cluster, since we miss observational point in the upstream region. It should be noted that, while the Coma relic is located in a sector where the group NGC\,4839 is falling in the main cluster, the E relic in A2345 is in a low-density region where no apparent accretion is currently ongoing. Therefore, a similarity between the two systems is not guaranteed. 

We did not attempt the modeling of the magnetic field profile in the W relic sector since an analytical description of the thermal electron density in this region is not trivial. Geometrical uncertainties could be the cause of the discrepancy between the observed MAD(RM) value of this relic and the model derived from the X-ray surface brightness profile (see Fig.~\ref{fig:RM_meanmodels}, bottom panel).

\section{Conclusions}
\label{sec:conclusion}

We investigated the intra-cluster magnetic field of the merging galaxy cluster Abell 2345 by using polarization observations of cluster radio sources and relics. We present new JVLA observations of this galaxy cluster in the 1-2 GHz L-band, with the angular resolution ranging from 3$\arcsec$ to 30.5$\arcsec$. These images reveal the complex internal structure of the two radio relics to the east (E relic) and to the west (W relic) of the cluster. In addition, we detected 5 sources seen in projection within a radius of 1 Mpc from the cluster center.

We applied RM synthesis and derived the average RM and its dispersion of each polarized source. We also analyzed a XMM-Newton archival observation which show a clearly disturbed morphology. The average RM radial profiles show a decreasing trend centered on the X-ray peak of the cluster, with the values obtained at the location of the most external source and of the relics being consistent with the Milky Way foreground. A decreasing trend is also observed as a function of the X-ray surface brightness.

We created 3D simulations of the galaxy cluster sector containing the E radio relic, including, both, a thermal electron density analytical profile derived from X-ray observations and a 3D magnetic field model based on MHD cosmological simulations \citep{Dominguez19}. We derived mock RM maps and compared the resulting RM median absolute deviation, MAD(RM), to observed values in order to constrain the parameters of the magnetic field model. This method relies on the assumption that all the observed polarized sources lie at the same distance along the line-of-sight and that the origin of the observed MAD(RM) decrease with the projected cluster radius is caused by the Faraday rotation in the ICM. 

We find that in our best model the magnetic field linearly decreases with the thermal electron density, with a power-law index $\eta=1$. This value is larger than the one obtained in cosmological simulations and for the Coma cluster, i.e. $\eta\sim0.5$ \citep{Bonafede10,Vazza18}. This implies that the magnetic field is not in equilibrium with the thermal gas. The best model has an average central magnetic field $\langle B_0 \rangle$ = $2.8\pm0.1 \ \mu$G while the average magnetic field at the position of the E relic is $\sim$ 0.3 $\mu$G. This value is $\sim$2.7 times lower than the equipartition estimates. The same model, derived for the E relic sector, is also able to describe the decrease of MAD(RM) with the X-ray surface brightness which is observed for all the sources in the cluster.

We compared our results with the literature, finding a good match, despite the variety of approaches used to obtain magnetic field estimates in galaxy clusters with different properties. Even with the large uncertainties that remains in the derivation of cluster magnetic field properties from RM data, a great improvement is constituted by the use of a realistic power spectrum derived from MHD cosmological simulations which was found to lead different results compared to a magnetic field model based on Kolmogorov power spectrum. In order to achieve a general understanding of the magnetic field structure (radial profile, power spectrum, connection to cluster properties) a larger number of this kind of studies should be performed. In particular, this is the first time that this analysis is performed using polarization and RM synthesis data of a cluster radio relic and more studies would help in confirming our findings.

\section*{Acknowledgements}
C.S. and A.B. acknowledge support from the ERC-StG DRANOEL, n. 714245. A.B. acknowledges support from the MIUR grant FARE ``SMS". C.S. acknowledges C. Haines for private communications about Arizona Cluster redshift survey (ACReS) results on Abell 2345, and P. Jagannathan for useful discussions about A-projection. L. L. acknowledges financial contribution from the contracts ASI-INAF Athena 2019-27-HH.0, ‘‘Attivit\`a di Studio per la comunit\`a scientifica di Astrofisica delle Alte Energie e Fisica Astroparticellare’’ (Accordo Attuativo ASI-INAF n. 2017-14-H.0), and from INAF ‘‘Call per interventi aggiuntivi a sostegno della ricerca di mainstream di INAF’’. F.V. and P.D.F. acknowledge support from the ERC-StG MAGCOW, n. 714196. We thank the anonymous referee whose comments contributed to the improvement of this paper. \newline

\section*{DATA AVAILABILITY}
The data underlying this article will be shared on reasonable request to the corresponding author.




\bibliographystyle{mnras}
\bibliography{my_bib} 







\bsp	
\label{lastpage}
\end{document}